\documentclass[preprint,3p,authoryear,a4paper]{elsarticle}

\usepackage{graphicx}
\usepackage[table]{xcolor}
\usepackage{dsfont}%%%%%%
\usepackage{amsmath,bbm,bm}
\usepackage{amssymb}%%%
\usepackage{amsfonts}
\usepackage{amsbsy}
\usepackage{amsthm}
\usepackage{array}
\usepackage{booktabs} % TBV
\usepackage{verbatim} % TBV
\usepackage[english]{babel}
\usepackage{float}
\usepackage{epsfig}
\usepackage{multirow}
\usepackage{epstopdf}
\usepackage{natbib}
\AtBeginDocument{}

\usepackage{enumitem}
\usepackage{booktabs, array}
\usepackage{lipsum, tabularx}
\usepackage{siunitx,lipsum}
\usepackage[linkbordercolor={1 0 0},citebordercolor={1 1 1}]{hyperref}
\usepackage{nameref}
\usepackage{color}
\usepackage{cancel}
\usepackage{lscape}
\usepackage{tabularx}
\usepackage{subfigure, ulem}
\usepackage{soul}
\usepackage[toc,page]{appendix}

\renewcommand\appendix{\par
	\setcounter{section}{0}%
	\setcounter{subsection}{0}%
	\setcounter{table}{0}
	\setcounter{table}{0}
	\setcounter{figure}{0}
	\gdef\thetable{\Alph{table}}
	\gdef\thefigure{\Alph{figure}}
	\gdef\thesection{\Alph{section}}
	\setcounter{section}{0}}

\newcommand\undermat[2]{
	\makebox[0pt][l]{$\smash{\underbrace{\phantom{%
					\begin{matrix}#2\end{matrix}}}_{\text{$#1$}}}$}#2}
				
\newcolumntype{L}[1]{>{\raggedright\let\newline\\\arraybackslash\hspace{0pt}}m{#1}}
\newcolumntype{C}[1]{>{\centering\let\newline\\\arraybackslash\hspace{0pt}}m{#1}}
\newcolumntype{R}[1]{>{\raggedleft\let\newline\\\arraybackslash\hspace{0pt}}m{#1}}

\renewcommand{\thefigure}{\arabic{section}.\arabic{figure}}

\renewcommand{\thetable}{\arabic{section}.\arabic{table}}

\newcommand{\AV}[1]{\textcolor{black}{ #1}}

\newcounter{arclist}

\newcounter{arcenum}

\makeatletter
\def\ps@pprintTitle{%
	\let\@oddhead\@empty
	\let\@evenhead\@empty
	\def\@oddfoot{}%
	\let\@evenfoot\@oddfoot}
\makeatother
\begin{document}
	
	\begin{frontmatter}
		
	\title{A multivariate evolutionary generalised linear model framework \\ with adaptive estimation for claims reserving}

	\author[UMelb]{Benjamin Avanzi}
	\ead{b.avanzi@unimelb.edu.au}
	
	\author[UNSW]{Greg Taylor}
	\ead{greg.taylor@unsw.edu.au}
	
	\author[UNSW,UdeM]{Phuong Anh Vu\corref{cor}}
	\ead{p.vu@unsw.edu.au}
	
	\author[UNSW]{Bernard Wong}
	\ead{bernard.wong@unsw.edu.au}
	
	\cortext[cor]{Corresponding author. }
	
\address[UMelb]{Centre for Actuarial Studies, Department of Economics \\ University of Melbourne VIC 3010, Australia}
	\address[UNSW]{School of Risk and Actuarial Studies, UNSW Australia Business School\\ UNSW Sydney NSW 2052, Australia}
		\address[UdeM]{D\'epartement de Math\'ematiques et de Statistique, Universit\'e de Montr\'eal \\ Montr\'eal QC  H3T 1J4, Canada}
	
		\begin{abstract}

		In this paper, we develop a multivariate evolutionary generalised linear model (GLM) framework for claims reserving, which allows for dynamic features of claims activity in conjunction with dependency across business lines to accurately assess claims reserves. We extend the traditional GLM reserving framework on two fronts: GLM fixed factors are allowed to evolve in a recursive manner, and dependence is incorporated in the specification of these factors using a common shock approach.
		
		We consider factors that evolve across accident years in conjunction with factors that evolve across calendar years. This two-dimensional evolution of factors is unconventional as a traditional evolutionary model typically considers the evolution in one single time dimension. This creates challenges for the estimation process, which we tackle in this paper. We develop the formulation of a particle filtering algorithm with parameter learning procedure. This is an adaptive estimation approach which updates evolving factors of the framework recursively over time.
		
We implement and illustrate our model with a simulated data set, as well as a set of real data from a Canadian insurer.

		\end{abstract}

		\begin{keyword}
		Claims Reserving, Evolutionary GLM, Adaptive Reserving, Particle Learning, Common Shock Models
			
			JEL code:
			G22  

MSC classes: 
%60G51 \sep % Processes with independent increments
%93E20 \sep % Optimal stochastic control
91G70 \sep 	%Statistical methods; risk measures [See also 62P05, 62P20] in Actuarial science and mathematical finance
91G60 \sep 	%Numerical methods (including Monte Carlo methods) in Actuarial science and mathematical finance
62P05 \sep 	%Applications of statistics to actuarial sciences and financial mathematics
62H12 %\sep 	%Estimation in multivariate analysis
%91B30 %\sep % Risk theory, insurance
		\end{keyword}

\newtheorem{remark}{Remark}[section]
\numberwithin{equation}{section}		
		
	\end{frontmatter}

\section{Introduction}\label{sec:Intro}
A general insurer (also known as a non-life insurer, or property and casualty insurer) usually experiences delays between the occurrences of insured events, their reporting and the actual payments of claims. Reserves for these outstanding claims are typically one of the largest liabilities on the balance sheet of the insurer, see for example, \citet*{AlWu09,SaGi14,AbBoCo15}. Therefore, they can have significant impacts on the emerging profit as well as solvency of the insurer. To accurately assess outstanding claims reserves, we argue in this paper that it is advantageous to both incorporate evolutionary \textit{and} dependence modelling in the process, and we develop a new methodology for doing so. The term ``evolutionary'' refers to the use of flexible modelling structures that adapt to changes in the claims experience. In the subsections below, we discuss the motivations for our developments in detail and further describe the contributions of our paper.\par 

\subsection{Evolutionary modelling in claims reserving}\label{sec:Evolmotiv}
\begin{figure}[H]
	\centering
	\includegraphics[scale=0.6]{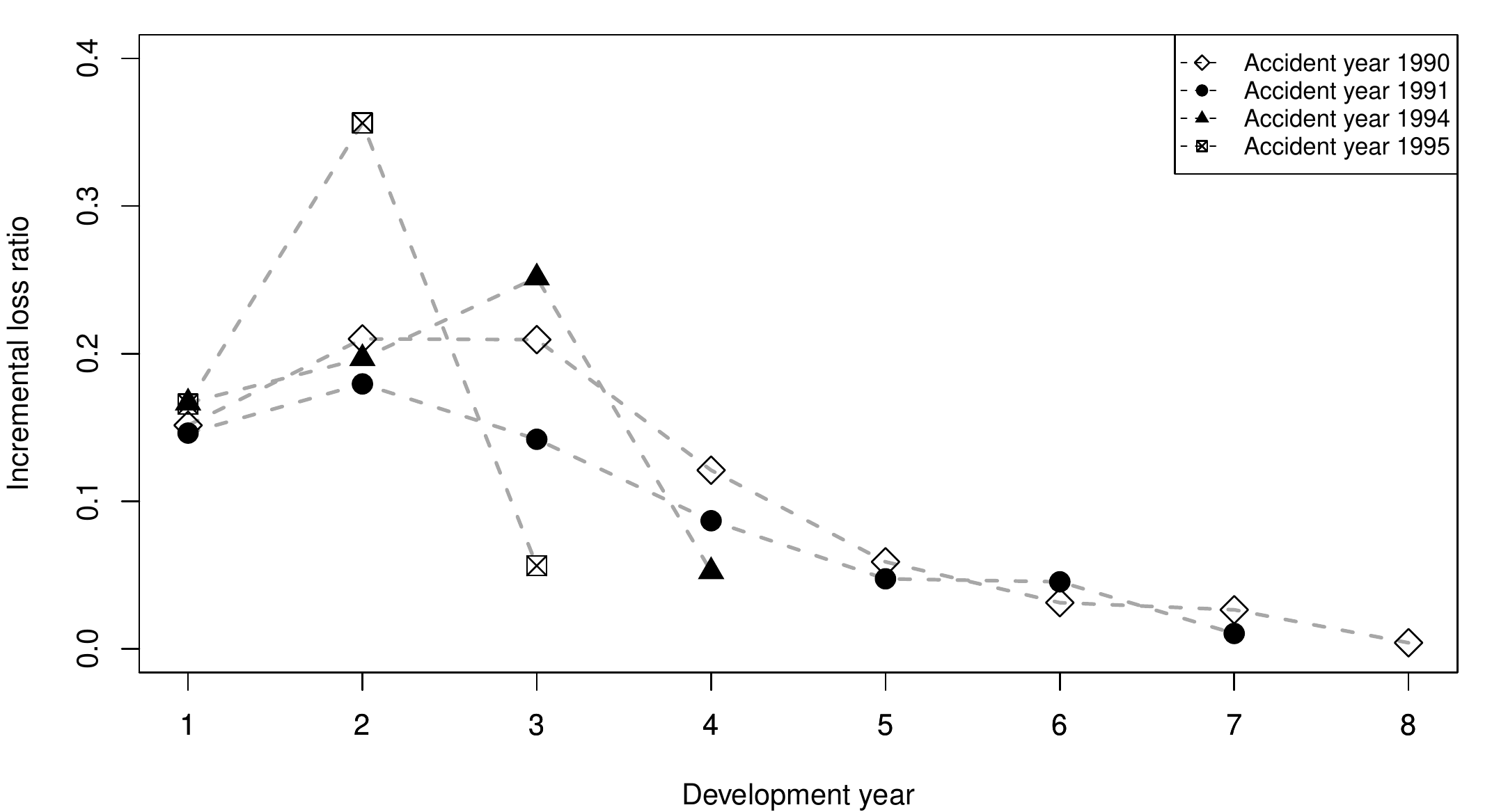}
	\caption{Incremental loss ratios in the Commercial Auto line of an American insurer (Schedule P)}\label{fig:scheduleP}
\end{figure}

The valuation of claims reserves is a predictive modelling activity where past information is used to predict outstanding claims for the future. It is, however, not at all unusual to observe changes in claim experience over time \citep*{Ghe01,Ren89,GlVe09}. These changes can be due to various reasons, such as legislative reforms, or changes to the  internal operations of insurers.  For example, the recent reform for Auto Bodily Injury covers in New South Wales (Australia) has resulted in faster claims resolution \citep*{SIRA18}, and has thus reduced payment delays. Another example is the development patterns from the Commercial Auto line of an American insurer whose data is used for illustration in \citet*{ShFr11}. Figure \ref{fig:scheduleP} provides plots of incremental loss ratios in four accident years 1990, 1991, 1994 and 1995 of this insurer. Incremental loss ratios are calculated as incremental claims divided by the total premium earned in the corresponding accident year. Changes in claim activity across accident years are evident in this figure with variations in the development patterns across accident years. \par 

When changes in development patterns have taken place, it is no longer straightforward to project future trends \citep*{Zeh94,Ghe01,Sim12}. Traditional reserving methods which use the assumption of similar claim activities across accident years, such as the traditional chain ladder algorithm and the traditional generalised linear model (GLM) framework, are no longer appropriate. In some cases, actuaries have to make a lot of judgements to remove or reduce the effects of these changes before applying traditional reserving methods. An example of this procedure is the Berquist-Sherman technique by \citet*{BeSh77}. These judgements are often difficult and time consuming to make, as well as to justify to management and peer reviewers \citep*{Sim12} because of their inherent subjective nature. In other cases, when the model selected using earlier data no longer fits more recent data, there may be a need to revise its algebraic structure. This will result in a fundamental discontinuity in the sequence of estimates such as the central estimates of outstanding claims \citep*{TaMcGr03}.  \par 

A very elegant and plausible solution to this issue is to accommodate changes directly in the model structures by allowing model factors to evolve \citep*{DeZe83,Zeh94,TaMcGr03, GlVe09}. These models have appeared in the literature under many names, including state space models \citep*{AlRi03,ChJo17}, adaptive models  \citep*{TaMc09}, robotic models \citep*{Mcg07}, and what we call them in this paper, \textit{evolutionary} models \citep*{Sim12}. These models not only can provide a better fit to the data, but also are more parsimonious. Trends in the data can be captured using a simple but explicit structure, enhancing model interpretation and reducing the need for unrecognised parameters \citep*{Zeh94}. \par 

\AV{Evolutionary models belong to a larger group known as mixed effect models where model factors are randomised, for example, \citet{Gus08,ZhDuGu12,ShBaMe12,GiPiSi13,GiPiSi19}. The key feature of evolutionary models that distinguishes \AV{them} from other mixed effect models is the evolution of \AV{the majority of} random effects, \AV{especially development year effect,} in a recursive manner. This evolution is typically specified using a time series structure. \AV{While some existing mixed effect models such as \citet{ZhDuGu12,ShBaMe12} use time series processes to capture the dynamics in one model aspect such as the residuals or calendar year effect, other factors are typically static. These fixed effects include development year factors with the underlying assumption of one single development pattern across all accident years. On the other hand, most factors in evolutionary models, especially development year and accident year effects, are allowed to evolve over time. Overall,} the development of observations is explained by a series of associated random factors. These models are also commonly known as state space models, a very powerful model class in time series analysis \citep{Har90}.}

\AV{Random factors in evolutionary models can be estimated in a traditional manner which utilises the entire history of data at once using methods such as maximum likelihood estimation, or Bayesian estimation.} \AV{It is worth noting that traditional Bayesian inference has been particularly favourable for mixed effect models such as \citet{ZhDuGu12,ShBaMe12}. However, as mentioned previously, most factors in evolutionary models are dynamic and the development of observations over time is explained by the associated random factors at that point in time. To optimise the benefits from this nature of evolutionary models, a special calibration method is typically used for these models known as adaptive estimation, \textit{``Bayesian filtering''}, or recursive Bayesian estimation \citep{DuKo12}. This is a special type of Bayesian inference. While traditional Bayesian inference uses the complete history of data for calibration of all factors at once, adaptive estimation recursively updates the estimates of random factors using new observations at each time point. Many benefits stem from this estimation technique.} In \AV{this} recursive Bayesian structure, more weight is given to recent data which improves the responsiveness of the models to changes \citep*{Zeh94,Tay00,AlRi03}. By \AV{estimating underlying factors recursively over time rather than all at once, changes are recognised gradually. As a result,} a clear picture of changes in the historical experience can be obtained, and estimates derived from these models are also smooth over time \citep*{TaMcGr03,Sim12}. Parameters are not estimated separately using scarce information for immature accident years but are projected recursively using the previous ones \citep*{GlVe09}. This also reduces the reliance on arbitrary modelling judgements \AV{which can be particularly beneficial when dealing with a small dataset.} \par 

Evolutionary models have appeared in the loss reserving literature since as early as the 1980s. These models can be categorised into two groups on the ground of distributional assumption used. The first group consists of Gaussian models which rely on the Gaussian assumption of observations and model factors. This is also the dominant group of evolutionary models in the literature. Examples of models in this group include \citet*{DeZe83,Ver89,Ver94,NtDe02,AtPiFe10,Dej06}. The popularity of the evolutionary Gaussian approach comes from its tractability in the estimation process. The recursive estimates of model factors are available in closed form through the use of a very popular procedure called the Kalman filter. The estimates of factors obtained from this filter are the best linear estimators in terms of the mean square error. Detailed descriptions of the Kalman filter as well its various applications can be found in the two comprehensive review books on evolutionary models (state space models) by \citet*{Har90} and \citet*{DuKo12}. \par 

The second group of evolutionary models is the group of non-Gaussian models. This is a remarkably smaller group. When the distributional assumptions deviate from the Gaussian assumption, the Kalman filter no longer provides the best linear estimates of factors in terms of mean square error. \citet*{TaMc09} developed a univariate evolutionary GLM  framework with Poisson and gamma errors. An estimation procedure called second-order Bayesian revision developed in \citet*{Tay08} is used to provide an approximate solution to this framework through the use of second-order Taylor expansions. \citet*{Sim11} then provided a simulation-based solution commonly known as a particle filter to the framework in \citet*{TaMc09}. \citet*{DoCh13} developed an evolutionary framework using the generalised-beta II family, and used conventional Bayesian inference for model estimation.  \par 

\subsection{Dependence modelling in claims reserving}\label{sec:evolutionarydepmotiv}
An insurer typically operates in multiple business segments, the number of which can be up to 100 in some cases \citep*{AvTaWo16}. These segments can be business lines or subsets of these. Dependency across different business segments is an important characteristic of claims for a typical general insurer. As defined in \citet*{AvTaWo16} using non-technical terms, it typically is the situation where the (departures from known trends of the) experience of one segment varies in sympathy with that of other segments. This experience can arise from many causes. At the very least, some segments share elements in their reporting procedure \citep*{ShFr11,Dej12} and hence any changes in the operational system can affect these segments simultaneously. Similarly, legislative changes can also have impacts on some segments such as the same business line in different geographical locations. There can also be claim causing events such as hailstorms that give rise to claims in multiple business lines (for example, motor lines and property lines).  \par 

While business segments are likely dependent to some extent, it is quite rare to observe cases of co-monotonicity \citep*{KiKeIs02}. Due to the lack of a perfectly positive dependence structure across segments, the volatility of claims on the aggregate portfolio level is reduced compared to the aggregation of volatility on the individual segment level. This reduction often is known as a ``diversification benefit" (\citealp*{ShFr11,Dej12,CoGeAb16,AvTaVuWo16}). Ignoring (or underestimating) this effect can result in an  over-estimation of risk margin and risk-based capital. Even if a certain degree of prudence is recommended, insurers should have as correct reserves as possible, not as large reserves as possible \citep*{Ajn94}. This is to ensure that capital is used parsimoniously while meeting solvency expectations \citep*{AvTaWo16}. Indeed, many insurance regulatory frameworks enable insurers to enjoy their diversification benefits in assessing the risk margins for their outstanding claims liabilities, as well as risk capital for their consolidated operations (\citealp*{AvTaWo16}).\par 

In many cases, the dependence across business segments arises from some calendar year factors that affect claims in the same calendar year within and across segments simultaneously \citep*{ShBaMe12,Dej12,Wut10}. For example, a legislative change in a particular calendar year can speed up the claims settlement processes in all business segments. A subset of the existing literature on multivariate reserving focuses on modelling calendar year dependence, see for example, \citet*{Wut10,Dej12,ShBaMe12,AbBoCo15}. \par 

Besides copulas which are popular dependence modelling tools in the reserving literature \citep*{ShFr11,Dej12,ZhDu13,AbBoCo15,CoGeAb16}, common shock approaches have also gained their popularity recently \citep*{Dej06,ShBaMe12,AvTaVuWo16,AvTaWo18}. The main advantage of these approaches stems from their ability to capture structural dependence arising from known relationships. In addition, they also allow some interpretability of the dependence structure as well as a parsimonious construction of correlation matrices of large dimensions \citep*{AvTaWo18}. 

\subsection{Contributions and outline of the paper} \label{S_contributions}
In this paper, we allow for dynamic features of claims activity \textit{in conjunction with} dependency across business lines to accurately assess claims reserves. To the best of our  knowledge, this is the first paper that combines both elements, and applies the methodology to real insurance data.

Specifically, we develop and illustrate a multivariate evolutionary GLM framework for claims reserving. It aims to extend the traditional GLM reserving framework on two fronts: GLM fixed factors are allowed to evolve in a recursive manner, and dependence is incorporated in the specification of these factors using a common shock approach.  This is to utilise the flexibility of the GLM structure while being able to enjoy various benefits of common shock approaches. We also formulate adaptive approaches that provide recursive real-time estimates of random factors in the evolutionary GLM framework. The ``evolutionarisation'' of the GLM fixed effects is a natural step. In the loss reserving literature, the exponential dispersion family (EDF) and its sub-class, the Tweedie family of distributions, have been used frequently in univariate as well as multivariate models. A comprehensive review of these models can be found in \citet*{TaMc16}. Models using the EDF are typically specified using the GLM framework which allows explanatory fixed factors to be incorporated in a flexible manner.

We consider factors that evolve across accident years in conjunction with factors that evolve across calendar years. This two-dimensional evolution of factors is unconventional as a traditional evolutionary model typically considers the evolution in one single time dimension. This creates challenges for the estimation process, which we tackle in this paper; see Section \ref{sec:evolutionarygenest}. 

The paper is organised as follows. Section \ref{sec:evolutionarymodel} introduces our model, a multivariate evolutionary GLM framework. An adaptive estimation procedure for this framework is developed in Section \ref{sec:evolutionaryest}. A simulation illustration is provided in Section \ref{sec:evolutionarysim}. Section \ref{sec:evolutionaryreal} provides an illustration using real data. We discuss those results, as well as some practical considerations, in Section \ref{S_Discussion}. Section \ref{sec:evolutionaryremarks} concludes the paper.

\section{A multivariate evolutionary GLM framework}\label{sec:evolutionarymodel}
In this section we develop a multivariate evolutionary GLM framework. Framework specifications are given in Section \ref{sec:evolutionaryspecs}. In the existing literature of evolutionary modelling, models are often presented in matrix form, see for example, the review books by \citet*{Har90,DuKo12}. This not only helps present the models more elegantly but is also useful in the presentation of model calibration. Therefore, we also introduce a matrix representation of our model in Section \ref{sec:evolutionarystatespace}. 

\subsection{Specifications}\label{sec:evolutionaryspecs}
The framework being introduced applies to loss triangles, the standard representation of loss reserving data \citep*{Tay00,WuMe08}. Consider a portfolio of $N$ lines (or segments) of business, each with a corresponding loss triangle. Each triangle contains a set of incremental claims $Y_{i,j}^{(n)}$ indexed by integers $i\, (i=1,...,I),$ $j\,(j=1,...,J)$ and $n\,(n=1,...,N)$ which represents the accident year, development year and line of business, respectively. For simplicity it is assumed that $I=J$. However, this assumption can be relaxed to give the general case of loss trapeziums. It also follows that the claim represented by $Y_{i,j}^{(n)}$ belongs to the calendar year $t=i+j-1\,(t=1,...,T)$. Note that irregular data shapes as well as missing values can be accommodated in a robust manner in this framework.
 
The multivariate evolutionary GLM framework consists of two components: an observation component and a state component. The observation component specifies the relation between observations and random factors. These random factors are not observed and are often referred to as states in the literature of evolutionary models \citep*{DuKo12}. The state component specifies the recursive evolution of random factors. These two components are described in more detail in the subsections below. 
 
\subsubsection{Observation component}
Each incremental claim $Y_{i,j}^{(n)}$ is specified in the observation component of the framework and may or may not be observed at the valuation date. It is specified in the same manner as the traditional GLM framework with the mean and variance such that
\begin{align}
E[Y_{i,j}^{(n)}] & = \mu_{i,j}^{(n)},\\
Var[Y_{i,j}^{(n)}] & = \phi^{(n)}\cdot V\left(\mu_{i,j}^{(n)}\right),
\end{align}
where $\mu_{i,j}^{(n)}$ is the mean parameter, $\phi^{(n)}$ is the dispersion parameter, and $V(.)$ is an appropriate variance function. When the distribution choice is restricted to a popular sub-family of the EDF, known as the Tweedie sub-family, the following variance function is specified with a power parameter $p$
\begin{align}
V\left(\mu_{i,j}^{(n)}\right) = \left(\mu_{i,j}^{(n)}\right)^p.
\end{align} 
A specific value of $p$ maps to a distribution from the Tweedie sub-family. Therefore, the Tweedie sub-family can be useful for modelling in the reserving context as it allows model uncertainty to be considered naturally through the estimation of the power parameter $p$ \citep*{AlWu09}.\par 
The mean structure (also known as the systematic component) relates the observations with a set of explanatory factors. In this framework, a modified Hoerl curve \AV{(a discrete version of a Gamma curve)} which allows for calendar year effect is used. Hoerl curves are used frequently in reserving, especially evolutionary reserving, see for example, \citet*{DeZe83,EnVe01,TaMc09}. \AV{A Hoerl curve is specified using a function of $j$ and $\log(j)$ which aims to approximate and provide smoothing for a claims development pattern}. Various benefits of a Hoerl curve include parsimonious modelling, robustness against fluctuations in observations, and extrapolation beyond the range of the observed development year. They are considered desirable in the evolutionary reserving context as they allow systematic changes in the claims experience over time. \par 

Using a log-link on a Hoerl curve, the mean structure is specified such that
\begin{align}
\log(\mu_{i,j}^{(n)}) = a_i^{(n)} + r_i^{(n)}\cdot \log(j) + s_i^{(n)}\cdot j + h_{t}^{(n)},\label{eq:meanstructure}
\end{align}
where $a_i^{(n)}$ is the accident year factor, $r_i^{(n)}$ and $s_i^{(n)}$ are factors of the Hoerl curve that specifies the development pattern for accident year $i$, and $h_t^{(n)}$ is the calendar year factor. Note that these ``factors" would often be called ``parameters" in static models but we choose to call them factors to avoid confusion with model parameters which do not evolve; see, e.g., Figure \ref{fig:structure}. It is worth emphasising that factors $a_i^{(n)},\,r_i^{(n)},\,s_i^{(n)}$ are accident-year-specific, whereas factor $h_t^{(n)}$ is calendar-year-specific, as noted by their subscripts. The above mean structure as well as the link function can be modified on a case-by-case basis.

A special case of the multivariate evolutionary GLM framework is a multivariate Gaussian model with Gaussian assumptions of observations (on which a log-transformation can be first applied if observations are log-normal). \AV{This model extends the existing Gaussian evolutionary models in the literature \citep{DeZe83,Ver89,NtDe02,AtPiFe10} by allowing for calendar year dependence across business lines.} The observation component in this case is specified such that
\begin{align}
Y_{i,j}^{(n)} = a_i^{(n)} + r_i^{(n)}\cdot \log(j) + s_i^{(n)}\cdot j + h_{t}^{(n)} + \varsigma_{i,j}^{(n)},\quad \varsigma_{i,j}^{(n)} \sim \text{Normal}\left(0,\sigma^2_{\varsigma^{(n)}}\right).
\end{align}

\begin{figure}[htb]
	\centering
	\includegraphics[scale=0.4]{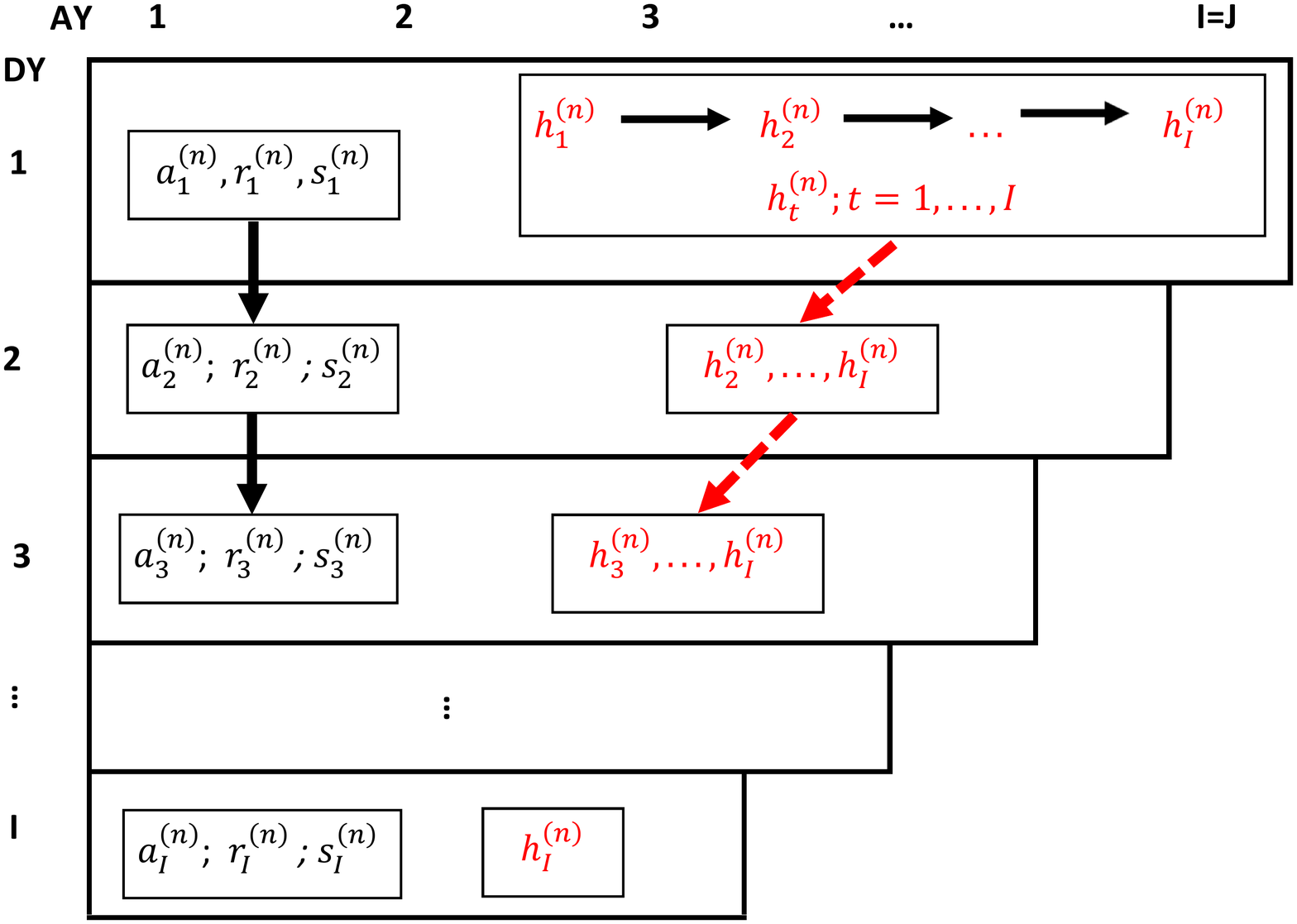}
	\caption{Evolution of random factors within a triangle (solid arrows represent evolution, dashed arrows represent one-to-one mapping, AY stands for accident year and DY stands for development year)}\label{fig:evolutionarydynamics} 
\end{figure}

\subsubsection{State component}\label{sec:states}
The observation component specified in the previous section is a standard GLM structure in reserving, see the review by \citet*{TaMc16}. \AV{In a standard structure, the same fixed (static) parameters are used to capture the development year effect (i.e $r_i^{(n)}=r^{(n)},\,s_i^{(n)}=s^{(n)}$ in a Hoerl curve specification) across all accident years. This essentially implies that one single average development pattern is assumed for all accident years.} The key features that differentiate the multivariate evolutionary GLM reserving framework from the traditional GLM reserving framework lie in the specification of the factors $a_i^{(n)},\,r_i^{(n)},\,s_i^{(n)}$ and $h_t^{(n)}$ of the mean structure. They are not unknown deterministic factors but are random and evolve over time. \AV{As a result, each accident year can have its own development pattern.} Their evolution can be specified using time series processes such as Autoregressive-Moving-Average (ARMA) processes. For simplicity, we use random walk processes for the evolution of states.

The evolution of $a_i^{(n)},\,r_i^{(n)},\,s_i^{(n)}$ is specified such that
\begin{align}
a_i^{(n)} & = a_{i-1}^{(n)} + {}_{a}\epsilon_i^{(n)}, \qquad {}_{a}\epsilon_i^{(n)} \sim \text{Normal}\left(0,\sigma^2_{{}_a\epsilon^{(n)}}\right),\\
r_i^{(n)} & = r_{i-1}^{(n)} + {}_{r}\epsilon_i^{(n)}, \qquad {}_{r}\epsilon_i^{(n)} \sim \text{Normal}\left(0,\sigma^2_{{}_r\epsilon^{(n)}}\right),\\
s_i^{(n)} & = s_{i-1}^{(n)} + {}_{s}\epsilon_i^{(n)}, \qquad {}_{s}\epsilon_i^{(n)} \sim \text{Normal}\left(0,\sigma^2_{{}_s\epsilon^{(n)}}\right),
\end{align}
where $\sigma^2_{{}_a\epsilon^{(n)}},\,\sigma^2_{{}_r\epsilon^{(n)}},\,\sigma^2_{{}_s\epsilon^{(n)}}$ are variances of the disturbance terms ${}_{a}\epsilon_i^{(n)},\,{}_{r}\epsilon_i^{(n)},\,{}_{s}\epsilon_i^{(n)}$, respectively, in the evolution. These variance terms are model parameters and need to be estimated. \par 

The evolution of calendar factor $h_t^{(n)}$ is specified using a modified random walk to incorporate calendar year dependence through a common shock approach
\begin{align}
h_t^{(n)}  & = h_{t-1}^{(n)} + {}_{h}\epsilon_t^{(n)}  + \lambda^{(n)}\cdot {}_h\tilde{\epsilon}_t, \qquad 
{}_{h}\epsilon_t^{(n)} \sim \text{Normal}\left(0,\sigma^2_{{}_h\epsilon^{(n)}}\right),\,
{}_h\tilde{\epsilon}_t \sim \text{Normal}\left(0,\sigma^2_{{}_h\tilde{\epsilon}}\right). \label{eq:calendarevolution}
\end{align}
There are two sources of disturbance in this evolution: line-specific disturbance ${}_{h}\epsilon_t^{(n)} $ and common shock disturbance ${}_h\tilde{\epsilon}_t$. The variances of these terms, $\sigma^2_{{}_h\epsilon^{(n)}}$ and $\sigma^2_{{}_h\tilde{\epsilon}}$, as well as the common shock coefficient $\lambda^{(n)}$ are model parameters. The calendar year dependence is \AV{induced} by the common shock term $ {}_h\tilde{\epsilon}_t$. This term can represent any changes in the calendar year $t$ that affect all lines simultaneously\AV{, such as superimposed inflation.} The effects of this common shock on each line, however, are usually not uniform as some lines can be more heavily affected than others. It is then desirable to use scaling factors $\lambda^{(n)}$ to adjust the effects of the common shock on individual lines. 
\begin{remark}
	\AV{In the specification of the framework, random walk processes are used for states evolution. This is chosen for the sake of simplicity following the existing literature \citep{DeZe83,Ver89,Ver94,AtPiFe10,TaMc09}. \AV{One also typically adheres to the simplest specification until there is reason to do otherwise. Further, the fact is that unpredictable nature of superimposed inflation in practice renders the random walk a reasonable representation. There can be cases where data can exhibit an accident year trend and/or calendar year trend\AV{, and in such specific cases} a different time series structure such as ARMA processes can be used. However, it is worth noting that} this comes at a cost of more extensive parametrisation.}
\end{remark}

The evolution of random factors in a single line can be summarised in Figure \ref{fig:evolutionarydynamics}. As demonstrated in this figure, we consider the evolution in the accident year dimension, meaning that random factors evolve and their estimates are updated as we proceed to the subsequent row of data in the loss triangles. This is to best utilise the available data in the first accident year to initialise the estimation process which will be discussed in Section \ref{sec:evolutionaryest}. It is worth noting that random factors $a_i^{(n)},\,r_i^{(n)},\,s_i^{(n)}$ are indexed by $i$ and they evolve from one accident year (i.e. row) to the next. Calendar factors $h_t^{(n)}$, on the other hand, evolve from one calendar year (i.e. diagonal) to the next and are indexed by $t$. However, when the process initiates at accident year 1, the effect of all calendar factors $h_1^{(n)},h_2^{(n)},...,h_I^{(n)}$ is present as claims in the first accident year are developed in these different calendar years.  Note that the relationship between these calendar year factors themselves follows the evolution specified in Equation \eqref{eq:calendarevolution}. These factors are mapped one to one with calendar factors in the next row $h_2^{(n)},..,h_I^{(n)}$ as we proceed to the second accident year, and so on. This is because claims within the same diagonal share the same calendar year effect.

\begin{figure}[H]
	\centering
	\includegraphics[scale=0.45]{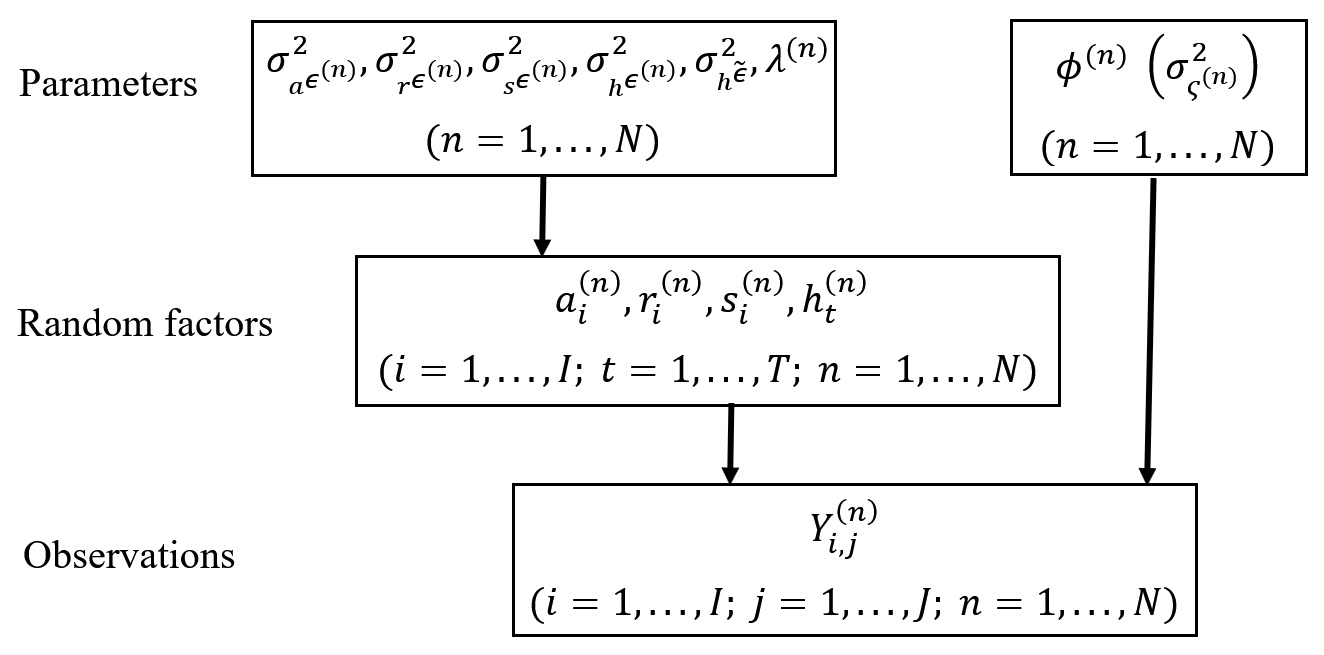}
	\caption{Summary of framework structure}\label{fig:structure}
\end{figure}
Figure \ref{fig:structure} summarises this section with the structure of the framework. There are three levels: underlying parameters, random factors and observations. Underlying parameters do not evolve with time and they specify the evolution of random factors as well as dynamics of the observations. Random factors evolve over time and they are explanatory factors of observations. Observations are directly observed from data and are specified by underlying parameters as well as random factors. \par 

\AV{As shown in Figure \ref{fig:structure}, there are $6N +1$ parameters, and $(3I + T) \times N$ random factors. \AV{This applies to a data set of $N$ triangles, each has size $I\times J$ (a total of $\frac{I\times (J+1)}{2}N$ observations).} The number of random factors is similar to that of a typical GLM framework which has a Hoerl curve specification with the addition of a calendar year factor, see for example, \citet{Zeh89,Wri90,EnVe02}. Similar models in which cross-classified structures are used instead of a Hoerl curve can be found in, for example, \citet{ShBaMe12,AbBoCo15}. However, in our framework, these random factors are specified to evolve recursively using previous random factors and some parameters. For immature accident years which have limited data, this structure can provide a better calibration of random factors for these years. }

\subsection{State space matrix representation}\label{sec:evolutionarystatespace}
In this section, we provide a matrix representation of the framework introduced in the previous section. \AV{A matrix representation of evolutionary models is often called a state space representation in the time series literature \citep{DuKo12}. It \AV{is a conventional representation of evolutionary models and }\AV{it} aims to provide a compact representation of the relationships between observations and random factors, and the development in random factors over time. Model estimation can then be presented more clearly using this representation. This is particularly useful in a multi-dimensional setting, which typically applies in the context of this paper when we have multiple observations at each time node and multiple underlying random factors.}
	
\subsubsection{Observation component}
In the proposed framework, loss reserving data can be considered as a multivariate time series process. Each accident year as a point in this process when new observations are assumed to arrive. The vector of observations at accident year $i$ is a vector of all $N\cdot(I-i+1)$ claims in the same accident year within and across triangles
\begin{align}
\boldsymbol{Y}_i = \begin{pmatrix}
Y_{i,1}^{(1)}\\
\vdots\\
Y_{i,I-i+1}^{(1)}\\
Y_{i,1}^{(2)}\\
\vdots\\
Y_{i,I-i+1}^{(N)}\\
\end{pmatrix},
\end{align} 
which has the following mean and dispersion 
\begin{align}
E[\boldsymbol{Y}_i] & = \boldsymbol{\mu}_i= \begin{pmatrix}
\mu_{i,1}^{(1)}\\
\vdots\\
\mu_{i,I-i+1}^{(1)}\\
\mu_{i,1}^{(2)}\\
\vdots\\
\mu_{i,I-i+1}^{(N)}\\
\end{pmatrix},\qquad \boldsymbol{\phi} = \begin{array}{c@{\!\!}l}
\left( \begin{array}{c}
\phi^{(1)}\\
\vdots\\
\phi^{(1)}\\
\phi^{(2)}\\
\vdots\\
\phi^{(N)}\\
\end{array}  \right)
&
\begin{array}{@{}l@{\,}l}
\left. \begin{array}{c} {} 
\\ {} \\ {} \end{array} \right\rbrace & \text{$(I-i+1)$ rows} \\
\begin{array}{c} {}  \\ {}
\\ {} \\ \end{array} & {} \\
\end{array}
\end{array}.
\end{align}
The mean structure is specified using a linear predictor with a log-link
\begin{align}
\log \left(\boldsymbol{\mu}_i\right) = \boldsymbol{A}_i\cdot\boldsymbol{\gamma}_i + \boldsymbol{E}_i \cdot\boldsymbol{\psi}_I, \label{eq:ssobservations}
\end{align}
where, with the mean structure specified in Equation \eqref{eq:meanstructure}, we have

\begin{align}
\boldsymbol{\gamma}_i^{(n)} & = \begin{pmatrix}
a_i^{(n)} \\
r_i^{(n)} \\
s_i^{(n)}\\
\end{pmatrix}, \qquad 
{\boldsymbol{\gamma}_i}  = \begin{pmatrix}
{\boldsymbol{\gamma}_i^{(1)}}\\
\vdots\\
{\boldsymbol{\gamma}_i^{(N)}}\\
\end{pmatrix}, \qquad
\boldsymbol{\psi}_I^{(n)}  = \begin{pmatrix}
h_1^{(n)} \\
\vdots \\
h_I^{(n)}\\
\end{pmatrix},\qquad \boldsymbol{\psi}_I  = \begin{pmatrix}
\boldsymbol{\psi}_I^{(1)}\\
\vdots\\
\boldsymbol{\psi}_I^{(N)}\\
\end{pmatrix},
\end{align}
\begin{align}
\boldsymbol{A}_i^{(n)} & = \begin{pmatrix}
1 & \log(1) & 1 \\
1 & \log(2) & 2 \\
\vdots & \vdots & \vdots\\
1 & \log(I-i+1) & I-i+1\\
\end{pmatrix}, \qquad 
\boldsymbol{A_i} = \begin{pmatrix}
\boldsymbol{A}_i^{(1)} & \boldsymbol{0} & \dots & \boldsymbol{0}\\
\boldsymbol{0} & \boldsymbol{A}_i^{(2)} & \dots & \boldsymbol{0}\\
\vdots & \vdots & \ddots & \vdots\\
\boldsymbol{0} & \boldsymbol{0} & \dots & \boldsymbol{A}_i^{(N)} \\
\end{pmatrix}, 
\end{align}
\begin{align}
\boldsymbol{E}_i^{(n)} & = \begin{array}{c@{\!\!}l}
\left( \begin{array}{ccccccc}
0 & \dots & 0 & 1 & 0 & \dots & 0 \\
0 & \dots & 0 & 0 & 1 & \dots & 0 \\
\vdots & \ddots & \vdots & \vdots & \vdots & \ddots & \vdots\\
\undermat{(i-1)$ cols$}{0 & \dots & 0} & \undermat{(I-i+1)$ cols$}{0 & 0 & \dots & 1} \\
\end{array}
\right)
&
\begin{array}{@{\!}l@{\!}l}
\left. \begin{array}{c} {} 
\\ {} \\ {} \\ {} \end{array} \right\rbrace & \text{\scriptsize{$(I-i+1)$ rows}} \\
\end{array}
\end{array}, \,
\boldsymbol{E}_i &= \begin{pmatrix}
\boldsymbol{E}_i^{(1)} & \boldsymbol{0} & \dots & \boldsymbol{0}\\
\boldsymbol{0} & \boldsymbol{E}_i^{(2)} & \dots & \boldsymbol{0}\\
\vdots & \vdots & \ddots & \vdots\\
\boldsymbol{0} & \boldsymbol{0} & \dots & \boldsymbol{E}_i^{(N)} \\
\end{pmatrix}.
\hspace{2cm}
\end{align}

In the special case of a multivariate Gaussian model, the observation equation can be specified in matrix form as
\begin{equation}
\boldsymbol{Y}_i = \boldsymbol{A}_i\cdot\boldsymbol{\gamma}_i+ \boldsymbol{E}_i\cdot \boldsymbol{\psi}_I + \boldsymbol{\varsigma}_i,\qquad \boldsymbol{\varsigma}_i {\sim}\text{Normal}(\boldsymbol{0},\boldsymbol{H}_i),
\end{equation}
where
\begin{align}
\boldsymbol{\varsigma}_i = \begin{pmatrix}
\varsigma_{i,1}^{(1)}\\
\vdots\\
\varsigma_{i,I-i+1}^{(1)}\\
\varsigma_{i,1}^{(2)}\\
\vdots\\
\varsigma_{i,I-i+1}^{(N)}\\
\end{pmatrix},
\end{align} 
\begin{align}
\boldsymbol{H}^{(n)}_i & = \left(\begin{array}{cccc}
\sigma^2_{\varsigma^{(n)}} & 0 & \dots & 0\\
0& \sigma^2_{\varsigma^{(n)}} & \dots & 0\\
\vdots & \vdots & \ddots & \vdots \\
\undermat{(I-i+1)$ cols$}{0 & 0 & \dots & \sigma^2_{\varsigma^{(n)}}}\\
\end{array}\right), \qquad \boldsymbol{H}_i = \begin{pmatrix}
\boldsymbol{H}^{(1)}_i & \boldsymbol{0} & \dots & \boldsymbol{0}\\
\boldsymbol{0} & \boldsymbol{H}^{(2)}_i & \dots & \boldsymbol{0}\\
\vdots & \vdots & \ddots & \vdots \\
\boldsymbol{0} & \boldsymbol{0}& \dots & \boldsymbol{H}^{(N)}_i \\
\end{pmatrix}.\\
\end{align}
\subsubsection{State component}
The random walk evolution of ${\boldsymbol{\gamma}_i}$ can be represented in matrix form as
\begin{equation}
{\boldsymbol{\gamma}_{i}} = \boldsymbol{\gamma}_{i-1} +\boldsymbol{{}_{\gamma}{\epsilon}}_{i},\qquad \boldsymbol{{}_\gamma{\epsilon}}_{i} \stackrel{}{\sim}\text{Normal}(\boldsymbol{0},\boldsymbol{Q}_{\boldsymbol{{}_\gamma\epsilon}}),\label{eq:ssstates}
\end{equation}
where 
\begin{align}
\boldsymbol{{}_{\gamma}\epsilon}_{i}^{(n)} & = \begin{pmatrix}
{}_{a}\epsilon_i^{(n)}\\
{}_{r}\epsilon_i^{(n)}\\
{}_{s}\epsilon_i^{(n)}\\
\end{pmatrix}, \qquad \boldsymbol{{}_{\gamma}{\epsilon}}_{i} = \begin{pmatrix}
\boldsymbol{{}_{\gamma}\epsilon}_{i}^{(1)}\\
\vdots\\
\boldsymbol{{}_{\gamma}\epsilon}_{i}^{(N)}\\
\end{pmatrix},\\
\boldsymbol{Q}^{(n)}_{\boldsymbol{{}_\gamma\epsilon}} & = \begin{pmatrix}
\sigma^2_{{}_a\epsilon^{(n)}} & 0 & 0\\
0& \sigma^2_{{}_r\epsilon^{(n)}} & 0\\
0 & 0 & \sigma^2_{{}_s\epsilon^{(n)}}\\
\end{pmatrix}, \qquad \boldsymbol{Q}_{\boldsymbol{{}_\gamma\epsilon}} = \begin{pmatrix}
\boldsymbol{Q}^{(1)}_{\boldsymbol{{}_\gamma\epsilon}} & \boldsymbol{0} & \dots & \boldsymbol{0}\\
\boldsymbol{0} & \boldsymbol{Q}^{(2)}_{\boldsymbol{{}_\gamma\epsilon}} & \dots & \boldsymbol{0}\\
\vdots & \vdots & \ddots & \vdots \\
\boldsymbol{0} & \boldsymbol{0}& \dots & \boldsymbol{Q}^{(N)}_{\boldsymbol{{}_\gamma\epsilon}} \\
\end{pmatrix}.
\end{align}

The evolution of calendar year factors is
\begin{align}
\boldsymbol{\psi}_t = \boldsymbol{R}_{t-1}\cdot\boldsymbol{\psi}_{t-1} + \boldsymbol{S}_{t-1}\cdot {}_h\boldsymbol{\epsilon}_t,\qquad {}_h\boldsymbol{\epsilon}_t \stackrel{}{\sim}\text{Normal}(\boldsymbol{0},\boldsymbol{Q}_{{}_h\boldsymbol{\epsilon}}) \label{eq:sscalendar}
\end{align}
where 
\begin{align}
\boldsymbol{\psi}_t^{(n)} & = \begin{pmatrix}
h_1^{(n)} \\
\vdots \\
h_t^{(n)}\\
\end{pmatrix},\qquad \boldsymbol{\psi}_t  = \begin{pmatrix}
\boldsymbol{\psi}_t^{(1)}\\
\vdots\\
\boldsymbol{\psi}_t^{(N)}\\
\end{pmatrix}, \qquad
{}_{h}\boldsymbol{\epsilon}_{t} = \begin{pmatrix}
{{}_{h}\epsilon_{t}^{(1)}}\\
\vdots\\
{{}_{h}\epsilon_{t}^{(N)}}\\
{{}_h\tilde{\epsilon}_{t}}\\
\end{pmatrix}, 
\end{align}
\begin{align}
\boldsymbol{R}_{t-1}^{(n)} &= \begin{array}{c@{\!\!}l}
\left( \begin{array}{cccc}
1 & 0 & \dots & 0  \\
0 & 1 & \dots & 0 \\
\vdots & \vdots & \ddots & \vdots \\
0 & 0 & \dots & 1 \\
\undermat{(t-1)$ cols$}{0 & 0 & \dots & 1} \\
\end{array}
\right)
&
\begin{array}{@{}l@{\,}l}
\left. \begin{array}{c} {} 
\\ {} \\ {} \\ {} \\ {} \end{array} \right\rbrace & \text{\small{$t$ rows}} \\
\end{array}
\end{array}, \qquad  \boldsymbol{R}_{t-1} = \begin{pmatrix}
\boldsymbol{R}_{t-1}^{(1)} & \boldsymbol{0} & \dots & \boldsymbol{0}\\
\boldsymbol{0} & \boldsymbol{R}_{t-1}^{(2)} & \dots & \boldsymbol{0}\\
\vdots & \vdots & \ddots & \vdots\\
\boldsymbol{0} & \boldsymbol{0} & \dots & \boldsymbol{R}_{t-1}^{(N)} \\
\end{pmatrix}, \hspace{3cm}\\
\boldsymbol{Q}_{{}_h\boldsymbol{\epsilon}} & = \begin{pmatrix}
\sigma^2_{{}_h\epsilon^{(1)}} & \dots & 0 & 0\\
\vdots & \ddots & \vdots & \vdots \\
0& \dots & \sigma^2_{{}_h\epsilon^{(N)}} & 0\\
0 &  \dots & 0& \sigma^2_{{}_h\tilde{\epsilon}}\\
\end{pmatrix},\qquad \boldsymbol{\Lambda}^{(n)} = \begin{array}{c@{\!\!}l}
\left( \begin{array}{c} 
0  \\
\vdots  \\
0 \\
\lambda^{(n)} \\
\end{array}
\right)
&
\begin{array}{@{}l@{\,}l}
\left. \begin{array}{c} {} 
\\ {} \\ {} \\ {}  \end{array} \right\rbrace & \text{\small{$t$ rows}} \\
\end{array}
\end{array},\\
\boldsymbol{S}_{t-1}^{(n)} & = \begin{array}{c@{\!\!}l}
\left( \begin{array}{c}
0  \\
\vdots  \\
0 \\
1 \\
\end{array}
\right)
&
\begin{array}{@{}l@{\,}l}
\left. \begin{array}{c} {} 
\\ {} \\ {} \\ {}  \end{array} \right\rbrace & \text{\small{$t$ rows}} \\
\end{array}
\end{array},\qquad 
\boldsymbol{S}_{t-1} = \begin{pmatrix}
\boldsymbol{S}_{t-1}^{(1)} & \boldsymbol{0} & \dots & \boldsymbol{0} & \boldsymbol{\Lambda}^{(1)} \\
\boldsymbol{0} & \boldsymbol{S}_{t-1}^{(2)} & \dots & \boldsymbol{0} & \boldsymbol{\Lambda}^{(2)} \\
\vdots & \vdots & \ddots & \vdots & \vdots \\
\boldsymbol{0} & \boldsymbol{0} & \dots & \boldsymbol{S}_{t-1}^{(N)} & \boldsymbol{\Lambda}^{(N)} \\
\end{pmatrix}. 
\end{align}
\section{Adaptive estimation}\label{sec:evolutionaryest}
This section covers the estimation of random factors and underlying parameters in the multivariate evolutionary GLM framework. The state space matrix representation of the framework is used in the development of the estimation approach. The random factors include non-calendar year factors ${\boldsymbol{\gamma}_i}$ ($i=1,\dots,I$) and calendar year factors $\boldsymbol{\psi}_I$. The unknown parameters are denoted as
\begin{align}
\boldsymbol{\Theta} = \lbrace \sigma^2_{{}_a\epsilon^{(n)}},\,\sigma^2_{{}_r\epsilon^{(n)}},\,\sigma^2_{{}_s\epsilon^{(n)}},\,\sigma^2_{{}_h\epsilon^{(n)}},\,\sigma^2_{{}_h\tilde{\epsilon}},\,\lambda^{(n)},\,{\phi}^{(n)};n=1,\dots,N \rbrace,
\end{align}
where ${\phi}^{(1)},\,\dots,\,\phi^{(N)}$ are replaced by $\sigma^2_{\varsigma^{(1)}},\,\dots,\,\sigma^2_{\varsigma^{(N)}}$ when a Gaussian model is considered.\par 

As mentioned in Section \ref{sec:Evolmotiv}, adaptive estimation is typically used for evolutionary models to estimate random factors recursively. This is in contrast with conventional estimation approaches that estimate model factors using all available information in loss triangles at once. Many benefits arise from this estimation approach \citep*{DuKo12}. \AV{The recursive Bayesian structure gives more weight to recent data which improves the responsiveness of model prediction to more recent experience. In addition, changes are recognised gradually over time, giving a clearer picture of changes in the historical experience. The recursive estimation and the framework structure in which random factors are specified recursively using previous factors and parameters also improve the calibration of random factors when data is scarce. However, it is worth noting that in order to have a good calibration of model parameters which feeds into the estimation of random factors, it is important to have sufficient data in the early periods of the estimation process. Hence, our calibration uses the accident year dimension as the main time dimension to utilise the greater availability of data in the early accident years. }

Motivated by \AV{the above} advantages, we adopt an adaptive estimation approach for our framework. A simulation based approach known as a particle filter with parameter learning approach is provided in Section \ref{sec:evolutionarygenest}. The adaptive estimates of random factors in the special case of Gaussian models are available in closed form and are also given in Section \ref{sec:evolutionarygausest}.  

\subsection{Particle filtering with parameter learning approach}\label{sec:evolutionarygenest}
In the adaptive estimation of evolutionary models, random factors are recursively estimated over time. Parameters, however, can be estimated in two ways: using off-line estimation methods and using on-line estimate methods \citep*{KaDoSiMa09}. Off-line estimation methods refer to the estimation of parameters using all observations at the end of the time series process. On-line estimation methods, on the other hand, integrate the estimation of parameters into the adaptive estimation of random factors. Hence parameters are estimated recursively in the same manner as random factors upon the arrival of new observations at each time point.  

The adaptive estimation approach chosen for the multivariate evolutionary GLM framework is an on-line estimation method known as a particle filter with parameter learning. It is based on the approach developed in \citet*{LiWe01}. This is a very popular method that has been used in various fields including traffic tracking \citep*{PoSo15}, medicine \citep*{LaRaAnLeEnSiMa18}, and finance \citep*{RiLo13}. Because it is an on-line estimation approach, parameters are updated recursively in the same manner as random factors even though they are assumed fixed effects. Artificial dynamics therefore are added to their specifications in a specific way. Because parameters are estimated in a Bayesian manner, their errors can also be assessed. The prior distribution of parameters at the initialisation can also be chosen to reflect the level of knowledge of these parameters. 

In this paper, we assume a time series process with accident years being the time dimension. However, as explained in Section \ref{S_contributions}, in the multivariate evolutionary GLM framework, while the random factors ${\boldsymbol{\gamma}_i}$ evolve by accident year, the factors $\boldsymbol{\psi}_I$---which is a vector of random calendar year factors---evolve by diagonals (see Figure \ref{fig:evolutionarydynamics}). To address this challenge, we treat calendar year factors in the same way as parameters in the adaptive estimation process. This is motivated by the fact that the effects of all calendar year factors are already present in the first accident year when the estimation process is initialised. \AV{These factors are not evolutionary across accident years, and as we proceed to the next accident year, we have more information about (some of) them, but their true value is not supposed to change.} Hence they can be treated as un-evolving factors beyond the first accident year and have the same nature as parameters in the framework. 

The notations used in the estimation process are from the state space matrix representation of the framework in Section \ref{sec:evolutionarystatespace}. Key equations are the observation equation \eqref{eq:ssobservations}, and the random factors equations \eqref{eq:ssstates} and \eqref{eq:sscalendar}. The estimation process is as follows.
\vspace{10pt}
\hrule
\noindent\textbf{Particle filter with parameter learning algorithm}
\vspace{2pt}
\hrule
\vspace{5pt}
\begin{enumerate}[label=\textbf{Step \arabic*}.,itemindent=*]
	\item Initialise.\par  
	At $i = 1$, for $m=1,...,M$, draw a sample (also called a particle) of parameters $\boldsymbol{\Theta}^{\{m;1\}}$ from their prior densities. The superscript $1$ is the filter time (accident year) index and the super script $m$ is the sample index.

	Draw calendar factors $\boldsymbol{\psi}_{I}^{\{m;1\}}$ using their specification in Equation \eqref{eq:calendarevolution} and $\boldsymbol{\Theta}^{\{m;1\}}$. The vector $\boldsymbol{\psi}_{I}^{\{m;1\}}$ represents the $m^{\text{th}}$ sample vector (particle) of all calendar factors (i.e. from calendar time 1 to calendar time $I$) at accident time 1.\par
	Draw initial samples of other factors at time 1
	\begin{align}
	{\boldsymbol{\gamma}_1^{\{m\}}} \sim {f}_{{\boldsymbol{\gamma}_{1}}}({\boldsymbol{\gamma}_{1}}; \boldsymbol{\Theta}^{\{m;1\}}).
	\end{align}
	Calculate the importance weights at time 1
	\begin{align}
	{\omega}^{\{m\}}_{1} = {f_{\boldsymbol{Y}_1|\boldsymbol{\gamma}_{1};\boldsymbol{\psi}_I,\boldsymbol{\Theta}}(\boldsymbol{y}_1|\boldsymbol{\gamma}_1^{\{m\}};\boldsymbol{\psi}_{I}^{\{m;1\}},{\boldsymbol{\Theta}}^{\{m;1\}})}.
	\end{align}
	This is to assign weight to each sample of simulated random factors and parameters using the likelihood of observations $\boldsymbol{Y}_1$ received at this time. \par 
	\hspace{-1cm}\underline{\textbf{For $i=2,...,I$ and $m=1,...,M$:}}
	\item Compute look-ahead (projected) random factors and parameters at time $i$ recursively using their values at $i-1$.\par
	Calendar year factors and parameters at time $i$ are projected using 
	\begin{align}
	\widetilde{\boldsymbol{\Theta}}^{\{m;i\}} & = \xi \cdot\boldsymbol{\Theta}^{\{m;i-1\}} + (1-\xi)\cdot\dfrac{1}{M}\cdot\sum_{{{r}}=1}^{M}\boldsymbol{\Theta}^{\{{{r};{i-1}}\}},\\
	\widetilde{\boldsymbol{\psi}}_{I}^{\{m;i\}} & = \xi \cdot \boldsymbol{\psi}_{I} ^{\{m;i-1\}} + (1-\xi)\cdot\dfrac{1}{M}\cdot\sum_{{{r}}=1}^{M}\boldsymbol{\psi}_{I}^{\{{{r};i-1}\}},
	\end{align}
	where $\widetilde{\boldsymbol{\Theta}}^{\{m;i\}}$ is the $m^{\text{th}}$ look-ahead estimate of $\boldsymbol{\Theta}$ at time $i$, $\xi$ is a shrinkage coefficient and $\widetilde{\boldsymbol{\psi}}_{I}^{\{m;i\}}$ represents the $m^{\text{th}}$ look-ahead sample of all calendar year factors at time $i$. \AV{\citet{LiWe01} discuss how to choose an appropriate value for $\xi$.}
	
	The $m^{\text{th}}$ look-ahead sample of non-calendar year factors at time $i$ is computed using
	\begin{align}
	\widetilde{\boldsymbol{\gamma}}_i^{\{m\}} = E[\boldsymbol{\gamma}_i|\boldsymbol{\gamma}_{i-1}^{\{m\}},\boldsymbol{\Theta}^{\{m;i-1\}}].
	\end{align}
	These samples are in the same nature as samples from prior distributions in the conventional Bayesian setting. \AV{Note that with a random walk formulation these are simply equal to their previous value at time $i-1$.}
	\item  Compute the look-ahead raw and normalised importance weights at time $i$.\par 
	Raw importance weights are calculated recursively using previous weights and updated likelihood of new observations $\boldsymbol{Y}_i$ received at $i$
	\begin{align}
	\tilde{\omega}^{\{m\}}_{i} = \omega_{i-1}^{\{m\}}\cdot f_{\boldsymbol{Y}_i|\boldsymbol{\gamma}_{i},\boldsymbol{\psi}_I;\boldsymbol{\Theta}}(\boldsymbol{y}_i|\widetilde{\boldsymbol{\gamma}}_i^{\{m\}},\widetilde{\boldsymbol{\psi}}_{I}^{\{m;i\}};\widetilde{\boldsymbol{\Theta}}^{\{m;i\}}),
	\end{align}
	where the likelihood is estimated using projected random factors and parameters.\par 
	Normalise the importance weights
	\begin{align}
	W_i^{\{m\}} = \dfrac{\tilde{\omega}_{i}^{\{m\}}}{\sum_{m=1}^{M}\tilde{\omega}_{i}^{\{m\}}}.
	\end{align}
	\item  Re-sample.\par 
	Resample $M$ samples (particles) of $\lbrace\boldsymbol{\gamma}_{i-1}^{\{m\}},\widetilde{\boldsymbol{\psi}}_{I}^{\{m;i\}};\widetilde{\boldsymbol{\Theta}}^{\{m;i\}}\rbrace_{m=1}^{M}$ with probabilities $\lbrace W_i^{\{m\}}\rbrace_{m=1}^{M}$.\par 
	Overall Steps 2-4 aim to select and resample particles (samples) of random factors and parameters. Samples of random factors and parameters that give higher likelihood on new observations $\boldsymbol{Y}_i$ received at $i$ will be sampled with higher probabilities.
	\item  Draw filtered (posterior) random factors and parameters.\par 
	Filtered parameters and calendar year factors are sampled using
	\begin{align}
	\boldsymbol{\Theta}^{\{m;i\}} & \sim \text{Normal}(\widetilde{\boldsymbol{\Theta}}^{\{m;i\}},(1-\xi^2){\boldsymbol{\Sigma}}_{\boldsymbol{\Theta}^{\{i-1\}}}),\\
	\boldsymbol{\psi}_{I}^{\{m;i\}} & \sim \text{Normal}(\widetilde{\boldsymbol{\psi}}_{I}^{\{m;i\}},(1-\xi^2){\boldsymbol{\Sigma}}_{\boldsymbol{\psi}_{I}^{\{i-1\}}}),
	\end{align}
	where ${\boldsymbol{\Theta}}^{\{m;i\}}$ is the $m^{\text{th}}$ filtered estimate of $\Theta$ at time $i$, ${\boldsymbol{\psi}}_{I}^{\{m;i\}}$ is the $m^{\text{th}}$ filtered sample of all calendar year factors at time $i$, ${\boldsymbol{\Sigma}}_{\boldsymbol{\Theta}^{\{i-1\}}}$ is the sample covariance matrix of $\lbrace{\boldsymbol{\Theta}}^{\{m;i-1\}}\rbrace_{m=1}^{M}$, and ${\boldsymbol{\Sigma}}_{\boldsymbol{\psi}_{I}^{\{i-1\}}}$ is the sample covariance matrix of $\lbrace{\boldsymbol{\psi}}_{I}^{\{m;i-1\}}\rbrace_{m=1}^{M}$. \par 
	Also draw samples of filtered values of non-calendar factors at time $i$ \AV{according to the distribution}
	\begin{align}
	\boldsymbol{\gamma}_i^{\{m\}} \sim {f}_{\boldsymbol{\gamma}_i|\boldsymbol{\gamma}_{i-1}}(\boldsymbol{\gamma}_i|\boldsymbol{\gamma}_{i-1}^{\{m\}};\boldsymbol{\Theta}^{\{m;i\}}).
	\end{align} 
	These filtered samples are different from the projected samples created in Step 2. Projected samples are drawn before observations $\boldsymbol{Y}_i$ are received at $i$. Filtered samples are drawn after the arrival of these observations. They are drawn directly using look-ahead samples which are resampled based on their likelihood on new observations. Hence filtered samples can be considered as correction samples, or posterior samples in the conventional Bayesian setting, of random factors and parameters for new observations. 
	\item  Calculate importance weights on filtered samples. \par 
	Importance weights are updated using 
	\begin{align}
	{\omega}^{\{m\}}_{i} = \dfrac{f_{\boldsymbol{Y}_i|\boldsymbol{\gamma}_{i},\boldsymbol{\psi}_I;\boldsymbol{\Theta}}(\boldsymbol{y}_i|\boldsymbol{\gamma}_i^{\{m\}},\boldsymbol{\psi}_{I}^{\{m;i\}};{\boldsymbol{\Theta}}^{\{m;i\}})}{f_{\boldsymbol{Y}_i|\boldsymbol{\gamma}_{i},\boldsymbol{\psi}_I;\boldsymbol{\Theta}}(\boldsymbol{y}_i|{\widetilde{\boldsymbol{\gamma}}}_i^{\{m\}},{\widetilde{\boldsymbol{\psi}}}_{I}^{\{m;i\}};\widetilde{\boldsymbol{\Theta}}^{\{m;i\}})}.
	\end{align}
	This update in weights take into account the mismatch between the likelihood at the actual sample of random factors $\boldsymbol{\gamma}_i^{\{m\}}$, calendar year factors $\boldsymbol{\psi}_I$, and parameters ${\boldsymbol{\Theta}}^{\{m;i\}}$, and their predicted values. Essentially samples that have higher predictive ability (with lower mismatches) are given higher weights and are more likely to be resampled in the next time period.
	\item  Repeat Steps 2-6 until $i=I$. 
\end{enumerate}
\hrule
\vspace{5pt}
\begin{remark}\label{re:filterlimitations}
	\AV{The filtering approach described in this section is a special type of particle filtering which incorporates parameter estimation. Particle filtering, in general, is considered an effective filtering tool for evolutionary models due to its flexibility in dealing with a wide range of models. This flexibility comes at a computational cost. However, with increasing computational power, it has found its applications in various fields \citep{PoSo15, LaRaAnLeEnSiMa18,RiLo13}.} \par 
	\AV{One often experience convergence issue when implementing a particle filter, see for example, \citet*{DoGoAn00,AnDoTa05,CaJoLoPo10,Cre12}. In the filtering approach developed in this paper, we have taken steps to assist convergence. In particular, while a traditional particle filter places the evaluation step ahead of the re-sampling step, this order is reversed in the filtering approach used. This in conjunction with somewhat informative priors/initial estimates used in the filter can reduce the degeneracy issue \citep{DoJo11,Cre12,CaGoMo07}. These also help reduce the computational cost of the filter and improve the rate of convergence.}
\end{remark}

\subsection{Dual Kalman filtering approach for Gaussian models}\label{sec:evolutionarygausest}
For Gaussian  models, a (modified) Kalman filter can be used to recursively estimate random factors. \AV{In the special case of Gaussian models, estimates of random factors can be obtained in analytical form, making it an interesting case to study.} As mentioned in the previous section, due to the calendar year factors $h_t$ which behave differently to other factors, the traditional Kalman filter cannot be applied without any adjustment. Using a similar treatment as in the previous section for the general case, we also consider calendar factors as static parameters to be updated beyond the first accident year. A modified version of the Kalman filter in the literature that fits well to this purpose is called the dual Kalman filter. It was developed by \citet*{NeSt76} and has been used to provide sequential estimates of dynamic factors as well as static factors or parameters of Gaussian models in various fields, including civil engineering \citep*{AzChPa15}, and vehicle systems \citep*{WeBuBlWi06}. The dual Kalman filter involves two filters that run in parallel, one for calendar factors, and one for non-calendar factors. The information from one filter flows into the other for continuing updates. Applying a dual Kalman filter to the Gaussian cases, we proceed as follows below. \newpage
\vspace{10pt}
\hrule
\noindent\textbf{Dual Kalman filter algorithm}
\vspace{2pt}
\hrule
\vspace{5pt}
\begin{enumerate}[label=\textbf{Step \arabic*}.,itemindent=*]
	\item Initialise.\par 
	At $i = 1$, obtain initial estimates of the mean and variance of calendar year factors
	\begin{align}
	\widetilde{\boldsymbol{\psi}}_{I}^{\{1\}} & = E[\boldsymbol{\psi}_{I}^{\{1\}} ],\\
	{}_h\widetilde{\boldsymbol{P}}_{1} & = Cov[\boldsymbol{\psi}_{I}^{\{1\}} ].
	\end{align} 
	These can be obtained by simulating $N$ samples of $\boldsymbol{\psi}_{I}^{\{1\}} $ using using their specification in Equation \eqref{eq:calendarevolution}. The mean and variance estimates can then be calculated using the sample mean and covariance matrix of these samples. \par 
	Also obtain initial estimates of non-calendar year factors
	\begin{align}
	\widetilde{\boldsymbol{\gamma}}_{1} &  = E[\boldsymbol{\gamma}_{1}],\\
	{}_\gamma\widetilde{\boldsymbol{P}}_{1} &= Cov[\boldsymbol{\gamma}_{1}],
	\end{align}
	which can be chosen using preliminary GLM analyses with fixed factors.\par
	\hspace{-1cm}\underline{\textbf{For $i=1,...,I$}}:
	\item Filter/update calendar year factors.\par 
	Calculate the Kalman gain ${}_h\boldsymbol{G}_i$ for calendar year factors 
	\begin{align}
	{}_h\boldsymbol{G}_i = {}_h\widetilde{\boldsymbol{P}}_{i}\cdot\boldsymbol{E}_i'\cdot\left(\boldsymbol{E}_i\cdot{}_h\widetilde{\boldsymbol{P}}_{i}\cdot\boldsymbol{E}_i' + \boldsymbol{H}_i\right)^{-1}.
	\end{align}
	Update estimates of calendar factors, including the conditional mean $\widehat{\boldsymbol{\psi}}_{I}^{\{i\}} $ and the conditional covariance matrix ${}_h\widehat{\boldsymbol{P}}_{i}$ on the arrival of observations $\boldsymbol{Y}_i$ at time $i$
	\begin{align}
	\widehat{\boldsymbol{\psi}}_{I}^{\{i\}} & = E[\boldsymbol{\psi}_{I}^{\{i\}}|\boldsymbol{Y}_i] = \widetilde{\boldsymbol{\psi}}_{I}^{\{i\}} + {}_h\boldsymbol{G}_i\cdot\left(\boldsymbol{Y}_i - \boldsymbol{A}_i\cdot\widehat{\boldsymbol{\gamma}}_{i-1} - \boldsymbol{E}_i\cdot\widetilde{\boldsymbol{\psi}}_{I}^{\{i\}} \right),\\
	{}_h\widehat{\boldsymbol{P}}_{i} & = Cov[\boldsymbol{\psi}_{I}^{\{i\}}|\boldsymbol{Y}_i] = {}_h\widetilde{\boldsymbol{P}}_{i} - {}_h\boldsymbol{G}_i\cdot\boldsymbol{E}_i\cdot{}_h\widetilde{\boldsymbol{P}}_{i}.
	\end{align}
 These can be considered as the posterior mean and variance in the conventional Bayesian setting.
	\item Filter/update non-calendar year factors.\par
	Calculate the Kalman gain ${}_{\gamma} \boldsymbol{G}_i$ for non-calendar year factors 
	\begin{align}
	{}_{\gamma} \boldsymbol{G}_i = {}_\gamma\widetilde{\boldsymbol{P}}_{i}\cdot\boldsymbol{A}_i'\cdot\left(\boldsymbol{A}_i\cdot{}_\gamma\widetilde{\boldsymbol{P}}_{i}\cdot\boldsymbol{A}_i' + \boldsymbol{H}_i\right)^{-1}.
	\end{align}
	Update estimates of other factors, including the conditional mean $\widehat{\boldsymbol{\gamma}}_{i}$ and the conditional covariance matrix ${}_\gamma\widehat{\boldsymbol{P}}_{i}$ using observations $\boldsymbol{Y}_i$ at time $i$
	\begin{align}
	\widehat{\boldsymbol{\gamma}}_{i} &= E[\boldsymbol{\gamma}_{i}|\boldsymbol{Y}_i] = \widetilde{\boldsymbol{\gamma}}_{i} + {}_{\gamma} \boldsymbol{G}_i\cdot\left(\boldsymbol{Y}_i - \boldsymbol{A}_i\cdot\widetilde{\boldsymbol{\gamma}}_{i} - \boldsymbol{E}_i\cdot\widehat{\boldsymbol{\psi}}_{I}^{\{i\}} \right),\\
	{}_\gamma\widehat{\boldsymbol{P}}_{i} &= Cov[\boldsymbol{\gamma}_{i}|\boldsymbol{Y}_i]  = {}_\gamma\widetilde{\boldsymbol{P}}_{i} - {}_{\gamma} \boldsymbol{G}_i\cdot\boldsymbol{A}_i\cdot{}_\gamma\widetilde{\boldsymbol{P}}_{i}.
	\end{align}
	These can be considered as the posterior mean and variance in the conventional Bayesian setting.
	\item Predict calendar year factors (time update).\par
	Project the calendar year factors in the next period
	\begin{align}
	\widetilde{\boldsymbol{\psi}}_{I}^{\{i+1\}} = E[\boldsymbol{\psi}_{I}^{\{i+1\}}|\boldsymbol{Y}_i] = \widehat{\boldsymbol{\psi}}_{I}^{\{i\}},
	\end{align}
	and project the error covariance of these factors
	\begin{align}
	{}_h\widetilde{\boldsymbol{P}}_{i+1} = Cov[\boldsymbol{\psi}_{I}^{\{i+1\}}|\boldsymbol{Y}_i] = {}_h\widehat{\boldsymbol{P}}_{i} + \boldsymbol{Q}_{{}_{h_I}\epsilon},
	\end{align}
	where $\boldsymbol{Q}_{{}_{h_I}\epsilon}$ is the artificial dynamic added to the covariance specification. It can be chosen to reflect the level of uncertainty regarding the estimates of calendar factors. Greater uncertainty can be accompanied by a larger artificial noise. \par 
	The estimates obtained in this step can be considered as prior mean and variance in the conventional Bayesian setting.
	\item  Predict non-calendar year factors (time update).\par
	Project non-calendar year factors ahead
	\begin{align}
	\widetilde{\boldsymbol{\gamma}}_{i+1} = E[\boldsymbol{\gamma}_{i+1}|\boldsymbol{Y}_i] = \widehat{\boldsymbol{\gamma}}_{i},
	\end{align}
	and project their error covariance ahead
	\begin{align}
	{}_\gamma\widetilde{\boldsymbol{P}}_{i+1} = Cov[\boldsymbol{\gamma}_{i+1}|\boldsymbol{Y}_i] = {}_\gamma\widehat{\boldsymbol{P}}_{i} + \boldsymbol{Q_{{}_{\gamma}\epsilon}}.
	\end{align}
	\item  Repeat step 2-5 until $i=I$.
\end{enumerate}
\hrule
\vspace{5pt}
\par The above algorithm is conditional on values of parameters $\boldsymbol{\Theta}$. Maximum likelihood estimation can be used to estimate these parameters. The log likelihood function can be written as
\begin{align}
&\log f_{\boldsymbol{Y}_{1:I}}(\boldsymbol{Y}_{1:I};\boldsymbol{\Theta})  = \sum_{i=1}^{I}\log f_{\boldsymbol{Y}_i|\boldsymbol{Y}_{i-1}}\left(\boldsymbol{y}_{i}|\boldsymbol{y}_{i-1};\boldsymbol{\Theta}\right)\\
& = -\dfrac{I^2(I+1)}{4}\cdot\log (2\pi)- \dfrac{1}{2} \sum_{i=1}^{I}\left(\log |\boldsymbol{A}_i\cdot{}_\gamma\widetilde{\boldsymbol{P}}_{i}\cdot\boldsymbol{A}_i'+\boldsymbol{E}_i\cdot{}_h\widetilde{\boldsymbol{P}}_{i}\cdot\boldsymbol{E}_i'+\boldsymbol{H}_i| + \right.\nonumber\\
&\left. \left(\boldsymbol{y}_{i}-\boldsymbol{A}_i\cdot\widetilde{\boldsymbol{\gamma}}_{i} - \boldsymbol{E}_i\cdot\widetilde{\boldsymbol{\psi}}_{I}^{\{i\}}\right)'\left(\boldsymbol{A}_i\cdot{}_\gamma\widetilde{\boldsymbol{P}}_{i}\cdot\boldsymbol{A}_i'+\boldsymbol{E}_i\cdot{}_h\widetilde{\boldsymbol{P}}_{i}\cdot\boldsymbol{E}_i'+\boldsymbol{H}_i\right)^{-1}\right.\nonumber\left.\times \left(\boldsymbol{y}_{i}-\boldsymbol{A}_i\cdot\widetilde{\boldsymbol{\gamma}}_{i} - \boldsymbol{E}_i\cdot\widetilde{\boldsymbol{\psi}}_{I}^{\{i\}}\right)\right),\nonumber
\end{align}
which can be maximised numerically to provide the maximum likelihood estimate of $\boldsymbol{\Theta}$. When maximum likelihood estimation is used for parameter estimation, bootstrapping is needed to assess the parameter uncertainty in the projection of claims liability due to a typically small sample size of loss data. \AV{This is because in that case the standard errors computed via maximum likelihood estimation are unreliable.}

\section{Simulation illustration}\label{sec:evolutionarysim}
A simulation illustration is performed for the evolutionary GLM framework to assess the performance of the estimation approach. The simulated data used for this illustration consists of two $15\times 15$ triangles given in Table \ref{tab:evolutionarysimdata21} and Table \ref{tab:evolutionarysimdata22} in Appendix \ref{sec:evolutionarysimdata2}. The data is simulated from a Tweedie GLM model with random factors having random walk evolution. The estimation approach developed in Section \ref{sec:evolutionarygenest} is used to estimate random factors and unknown parameters. We use 50,000 samples for each time period and initialise the estimation using static GLM estimates.

\subsection{Random factors estimation}
\begin{figure}[H]
	\centering
	\includegraphics[scale=0.7]{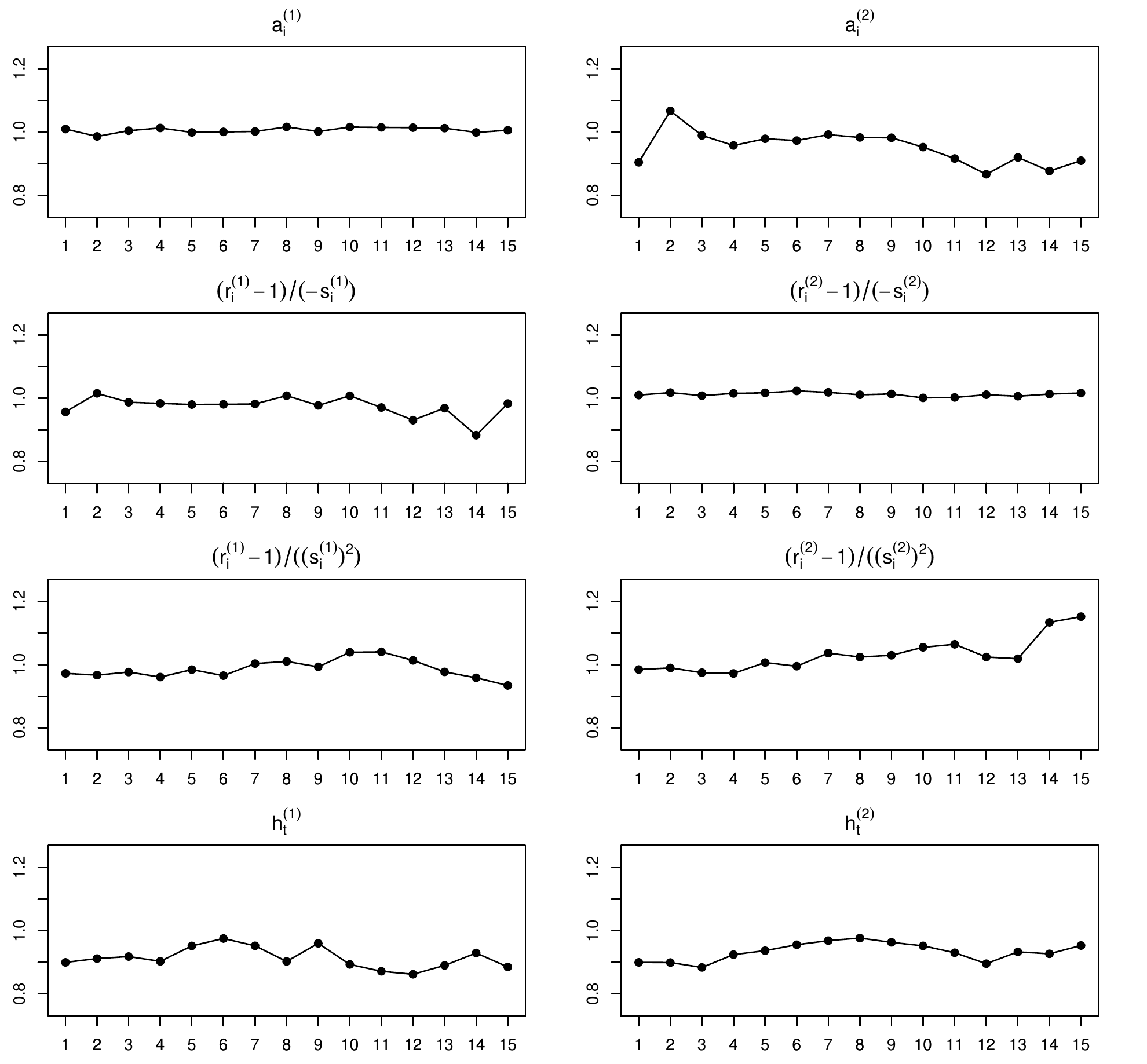}
	\caption{Fitted ratios of random factors in the GLM framework illustration (filtered values to true values)}\label{fig:randomfitted2} 
\end{figure}

Estimates of random factors are provided in Table \ref{tab:randomestsim2} in Appendix \ref{sec:evolutionarysimdata2}. Note that these are filtered estimates, or posterior estimates, meaning that the estimates  of factors in an accident period are obtained as the data in that accident period arrives. We provide plots of fitting ratios in Figure \ref{fig:randomfitted2}, which are calculated as ratios of filtered to true values. It is noted that each development pattern is fitted with a Hoerl curve orchestrated by two random factors $r_i^{(n)}$ and $s_i^{(n)}$. Therefore, the most direct way to assess the goodness-of-fit for the development pattern is to consider the fitting ratios of the mean and variance of the Hoerl curve calculated using these two parameters. In particular, the mean of the Hoerl curve for {accident period} $i$ and line $n$ is calculated by 
\begin{equation}
\dfrac{r_i^{(n)}-1}{-s_i^{(n)}},
\end{equation}
and the variance by 
\begin{align}
\dfrac{r_i^{(n)}-1}{\left(s_i^{(n)}\right)^2}.
\end{align}
It can be observed from Figure \ref{fig:randomfitted2} that the goodness of fit is reasonable for accident year factors $a_i^{(n)}$. The estimates of the means and variances of the Hoerl curves are also quite close to their actual values. Estimates of calendar factor $h_t^{(n)}$ are consistently lower than their true values, however the deviations remain within 20\% of actual values. \par 

A compensation across random factors is evident from their estimates in Table \ref{tab:randomestsim2}. Accident year factors $a_i^{(n)}$ are overestimated, and this is compensated by an underestimation of calendar year factors. A consistent mis-allocation between these factors is observed from relatively stable fitting ratios of these factors in Figure \ref{fig:randomfitted2}. This issue is not only specific to evolutionary models but is also very common in models that consider all three factors: accident year factor, calendar year factor and development year factor \citep*{Zeh94,Tay00,BaZe00,BrVe09,GlVe09,VeGuGa19}. Due to the collinearity between these factors, mis-allocations between them can largely offset to give an overall reasonable fit. Experienced practitioners also tend to be more interested in the combination of these factors rather than their individual trends \citep*{VeGuGa19}. As the ultimate goal of any valuation task is to forecast outstanding claims, a question is then raised regarding the impacts of this mis-allocation on the accurateness of claims projection. \citet*{McTaMi18} argued that extrapolating future trends of these factors using their corresponding estimates from the past would produce reasonably accurate future claims experience. This works well for cases with constant calendar year trends. When the trends are not constant, one should proceed with caution and some reasonableness checks of the trends can be useful.

To assess the calendar year dependence driven by common shocks, we calculate the Pearson correlation coefficient between actual calendar year factors $h_t^{(1)}$ and $h_t^{(2)}$, and the correlation coefficient between the estimates of these factors obtained from the particle filter. The Pearson correlation coefficient between the filtered (posterior) values is 0.4132 with a 95\% confidence interval (-0.1257; 0.7638). The actual Pearson correlation coefficient is 0.1463 which lies well within this interval. 

\subsection{Parameter estimation}
Parameter estimates are provided in Table \ref{tab:evolutionarysimest2} in Appendix \ref{sec:evolutionarysimdata2}. The results show that the true values of many parameters fall out of their respective confidence intervals, even though they appear close to their estimates.\par 

The narrow CIs that we observe in the results are mainly due to the degeneracy issue of the particle filter used. Particle degeneracy refers to the situation where all but a few particles (samples) have negligible weights. As a result, the re-sampling step in the particle filtering with parameter learning algorithm focuses on multiplying the few particles with significant weights and abandons the majority of particles with negligible weights. The resultant sample then has a very low diversity of particles. This consequently results in smaller confidence intervals of parameter estimates in the particle filtering with parameter learning \citep*{RiLo13}. \AV{As a result, the algorithm may not converge to the right parameters.} \AV{This issue has been recognised for particle filtering with parameter learning specifically (see for example, \citealp*{CaJoLo10,RiLo13}), and for simulation-based estimation approaches in general (see for example, \citealp*{DoGoAn00,AnDoTa05,CaJoLoPo10,Cre12}).} This illustration also has a large vector of observations at each time period (up to 30 in the first year). This high dimension of observations makes the likelihood function very steep and it further contributes to the degeneracy problem (see, for example, \citealp*{WaLiSuCo17,LiSuSaCo14}). \par 

\AV{Using informative prior distributions for parameters and initial values for random factors can help {reduce} this problem. In this illustration,  we have used estimates from a static GLM fitting as a guidance for selecting prior distributions and initial values for random factors. It is also worth noting that while it is desirable for prior distributions of parameters to have large variances, these prior distributions should not be too vague to avoid particle degeneracy issues. We have also performed a number of trial runs to get a reasonable particle filtering path which does not suffer a severe particle degeneracy issue. This task is quite similar in nature to the ``tuning" process of the Metropolis-Hastings algorithm in Bayesian inference. However, as mentioned in Remark \ref{re:filterlimitations}, the degeneracy issue is unavoidable in particle filtering and it is only possible to reduce it to a certain extent but not eliminate it completely.} \par 

As shown in Table \ref{tab:evolutionarysimest2}, there appears to be some compensation between variance terms in this illustration as well. The dispersion parameters of the observations, including $	\phi^{(1)},\,\phi^{(2)},\,p^{(1)},\,p^{(2)}$ are overestimated, while variance terms of random factors are underestimated. An explanation can be that while the framework has two components: observations and random factors, random factors are latent and only the total volatility is observed. This may cause a mis-allocation across variance terms. This issue is unlikely to distort the projection of claims if it is handled with caution. For example, observed rapid changes in claims development patterns should be appropriately recognised with large variance estimates of development factors. As long as the overall goodness-of-fit is reasonable, the compensation across component volatilities should not have material impacts on the projection of future claims because the total volatility of future claims is the subject of interest. A careful examination of data features as well as any expert opinions can help select more informative starting values and prior distributions of variance parameters. \AV{As mentioned above, in this illustration, we have based our initial estimates on static GLM estimates and performed trial runs to select appropriate prior distributions for variance parameters.}

\subsection{Goodness-of-fit analysis}
\begin{figure}[H]
	\centering
	\includegraphics[scale=0.7]{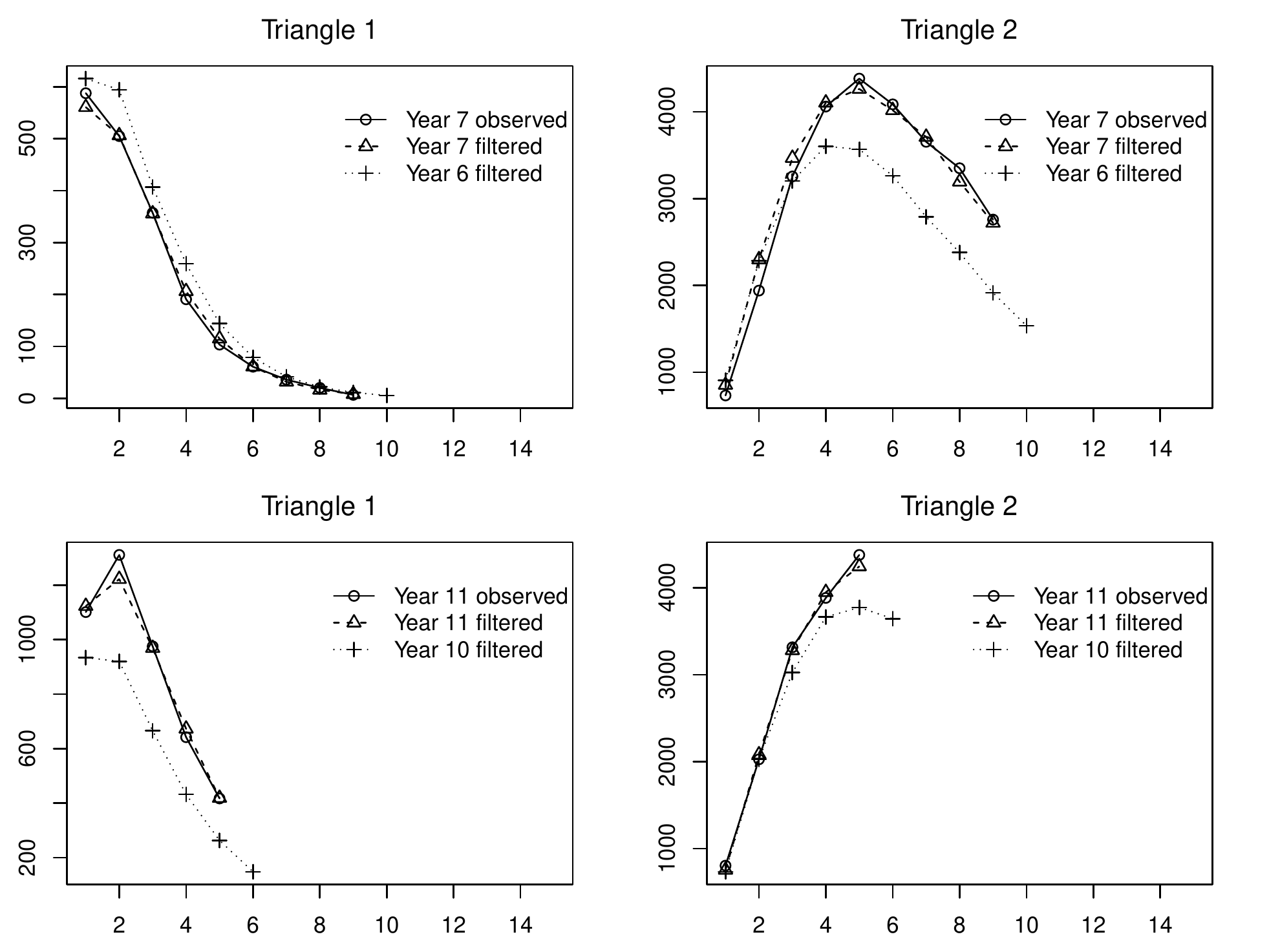}
	\caption{Tracking of claims development patterns for some accident years in the simulation illustration}\label{fig:glmtracking}
\end{figure}

We examine the performance of the particle filter by assessing how closely the estimated claim patterns track the observed patterns. Examples of this tracking for accident years 7 and 11 are given in Figure \ref{fig:glmtracking}. There are significant changes in the claims development patterns from year 6 to year 7, and from year 10 to year 11, as shown in this figure. However, the particle filter is able to track these changes very closely. \par 

\section{Real data illustration}\label{sec:evolutionaryreal}
The data set used for this illustration is from \citet*{CoGeAb16}. The two triangles chosen for illustration in this chapter are from the Accident Benefits covers of the Auto Insurance line in Ontario. One triangle is for Accident Benefits (AB) excluding Disability Income, and the other is for AB with Disability Income only. Incremental losses are given in Table \ref{Tab:evolutionaryreal1} and \ref{Tab:evolutionaryreal2} in Appendix \ref{sec:evolutionaryrealdata}. Claims are standardised using the total premium earned in the corresponding accident years to give incremental loss ratios.

\begin{figure}[htb]
	\centering
	\includegraphics[scale=0.5]{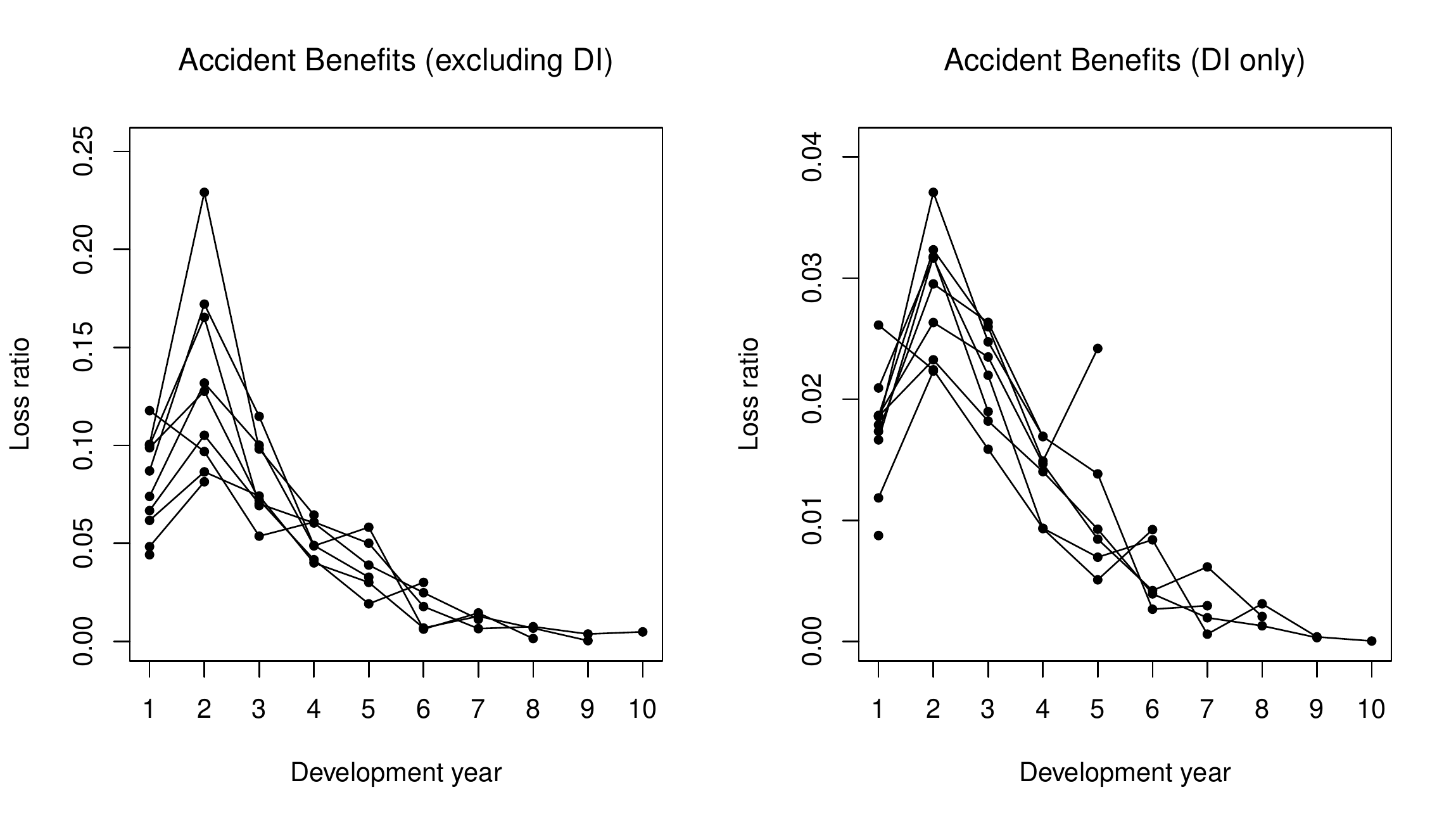}
	\caption{Incremental loss ratios in real data from a Canadian insurer}\label{Fig:evolutionarydevpatterns}
\end{figure} 

\subsection{Preliminary analysis}
A preliminary analysis is performed to assess the suitability of the data set. This includes an assessment of any changes in {the} development patterns as well as the dependence {across the two lines}. 
\subsubsection{Analysis of development patterns}
\begin{figure}[htb]
	\centering
	\includegraphics[scale=0.5]{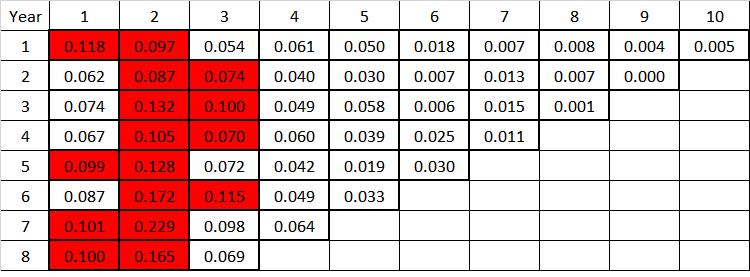}
	\centering
	\includegraphics[scale=0.5]{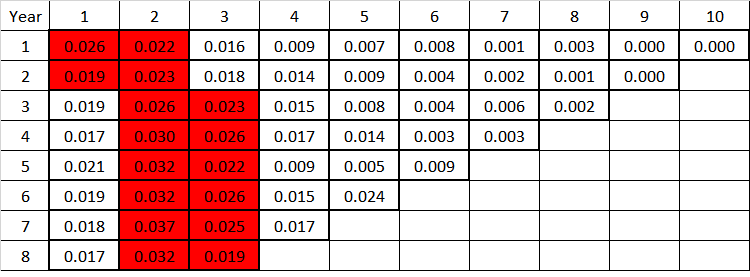}
	\caption{Loss triangles with top two values highlighted for each accident year (top: Accident Benefits excluding Disability Income, bottom: Accident Benefits - Disability Income only)}\label{fig:evolutionarypeaks}
\end{figure}

Plots of loss ratios are provided in Figure \ref{Fig:evolutionarydevpatterns}. For each loss triangle, the top two values in each accident year are also highlighted to identify the peak in the development pattern. These are provided in Figure \ref{fig:evolutionarypeaks} for accident years 1-8. \par  

Plots of loss ratios show variations in claims development patterns over time. In the Accident Benefits excluding Disability Income {line}, the peak in the claims development pattern shifts across development years 1-2 and 2-3. Hence it is desirable for the model to be able to capture this feature. In the Accident Benefits with Disability Income only line, the peak in the development shifts from development years 1-2 in the first two accident years to years 2-3 onwards. It could be tempting to model the first two accident years separately using a static modelling approach. However, using an evolutionary model can allow the changes in the prediction of claims to be smoothed out over time. In addition, it can be observed that the level of variation between claims in development years 1 and 3 within the same accident year has also varied over time. This indicates some variation in the development patterns besides the peaks. 

\subsubsection{Exploratory dependence analysis}\label{sec:exploratoryanalysis}
A heuristic dependence analysis is performed by fitting to each line a Tweedie GLM. \AV{The Tweedie family is a major subclass of the Exponential Dispersion Family consisting of symmetric and non-symmetric, light-tailed and heavy-tailed distributions \citep{AlLaSh152,Jor97}. This subclass includes many commonly used distributions in the reserving literature, see for example, \citet{AlWu09,BoDa11,EnVe02,PeShWu09}. A key feature of this subclass is the relationship between the unit variance function $V(\mu)$ and the mean $\mu$ where $V(\mu) = \mu^p$. The power parameter $p$ identifies a member of the family, for example, $p=1$ corresponds to a Poisson distribution, and $p=2$ corresponds to a gamma distribution. Through the flexibility of the power parameter $p$, model error is also allowed for.}\par 
\AV{The log-link and a modified Hoerl curve is used in this GLM structure} with additional covariates for the first two development years 
\begin{equation}
a_i^{(n)} + r^{(n)}\log(j) + s^{(n)}j + {b_1^{(n)}\mathbbm{1}_{\lbrace j=1\rbrace} + b_2^{(n)}\mathbbm{1}_{\lbrace j=2\rbrace}}.
\end{equation}
As the peak in the development pattern shifts across development years 1, 2 and 3, adding the two covariates for the first two development years can improve the goodness-of-fit of the Hoerl curve. Such modification is quite common in the applications of the Hoerl curve \citep*{EnVe01}. This static GLM fitting aims to remove fixed accident year and development year effects.\par 
\begin{table}[htb]
	\centering
	\begin{tabular}{ccc}
		\toprule
		\hline
		Pearson & Spearman & Kendall \\ 
		\hline
		0.2599 (0.0554) & 0.3087 (0.0222)& 0.2256 (0.0150) \\ 
		\hline
		\bottomrule
	\end{tabular}
	\caption{Measures of association between cell-wise GLM residuals and their corresponding $p$-values}\label{Tab:evolutionarycorcalca}
\end{table}
Measures of association between pair-wise GLM Pearson residuals of the two lines are provided in Table \ref{Tab:evolutionarycorcalca}. All coefficients are quite strong, \AV{and while the Pearson coefficient is not significant at 5\%, more holistic indicators such as Spearman and Kendall are significant.} Another GLM analysis is also performed with the chain ladder mean structure. The Pearson correlation coefficient of residuals from this GLM analysis, however, is significant. This further illustrates the conclusion in \citet*{AvTaWo16} that correlation coefficients are dependent on the methodology used. From these results, it is then reasonable to conclude that there may be dependence retained in the data set after removing fixed accident year and development year effects. \AV{However, this dependence is not necessarily linear, and a Pearson correlation coefficient may not be an appropriate model for it.} We also analyse heat maps of GLM Pearson residuals provided in Figure \ref{fig:evolutionaryheatcal} in Appendix \ref{sec:evolutionaryrealdata} and observe some common patterns in the diagonals of the two triangles. \par

\begin{figure}[H]
	\centering
	\includegraphics[scale=0.5]{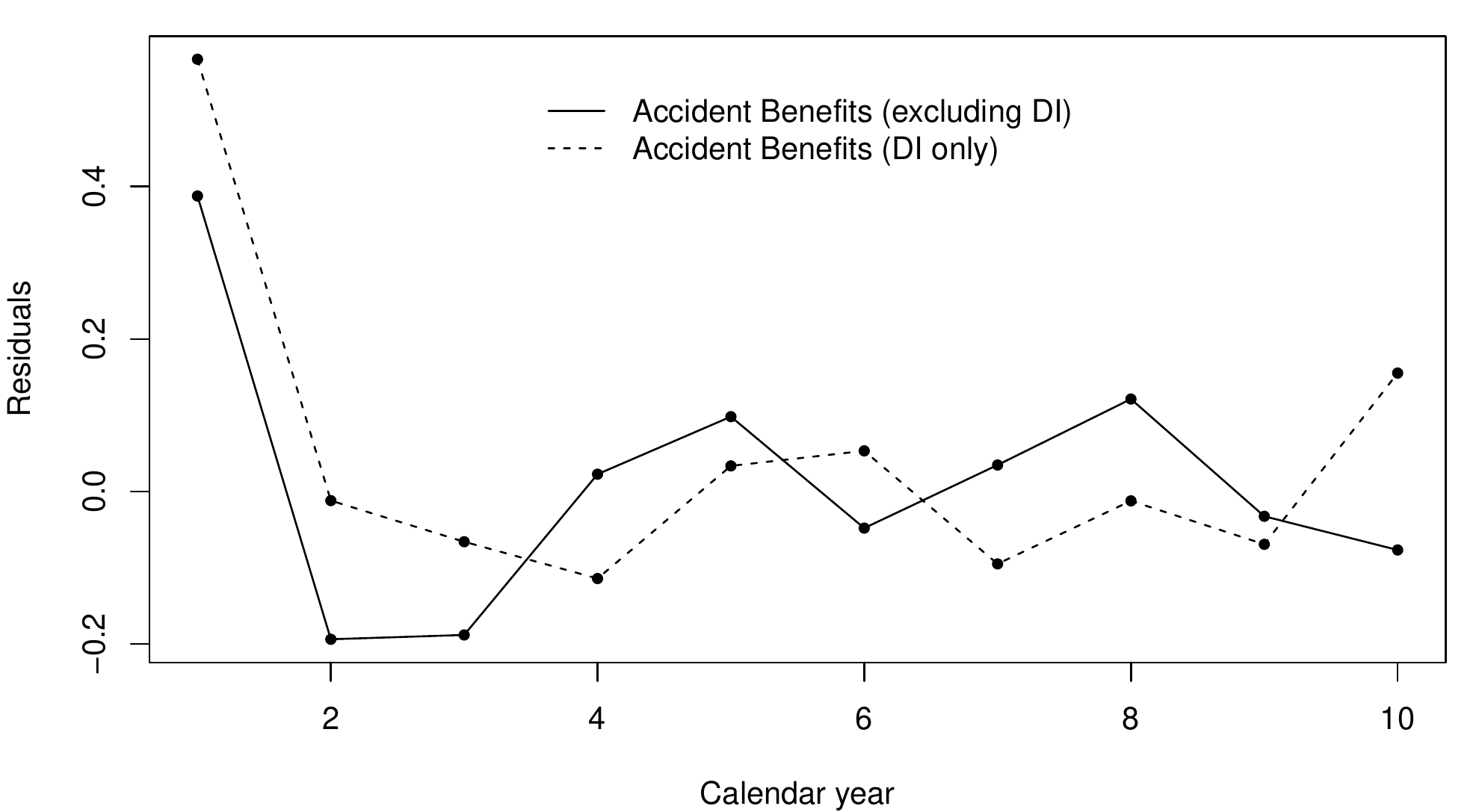}
	\caption{Plots of GLM residuals by calendar years}\label{fig:evolutionarycalres}
\end{figure}

We now examine residuals by calendar years to find further trace of calendar year trend and dependence. The residual for each calendar year is calculated as the difference between the sum of observed values and the sum of fitted values for all cells in that year, standardised using the sum of fitted values. Plots of calendar year residuals for the two lines are given in Figure \ref{fig:evolutionarycalres}. Clear evidence of calendar year dependence is observed from this figure. In particular, there are some sympathetic movements in the calendar year trends from both lines such as in calendar years 1 to 2, 4 to 5, 7 to 9. This suggests some common effects that impact both lines, as well as idiosyncratic effects within individual lines. The Pearson correlation coefficient between the calendar year residuals from the two lines is 0.6976 ($p$-value 0.0249), which is strong and significant. However, the correlation is -0.0691(p-value 0.8598) if CY1 is omitted. This highlights the limitations of using correlation to model dependence in this context; this is further discussed in Section \ref{S_limitcorr}.

The exploratory dependence analysis shows evidence of calendar year dependence across the two lines. Hence this data set is suitable for illustration of the multivariate evolutionary GLM framework.

\begin{figure}[htb]
	\centering     
	\includegraphics[scale=.6]{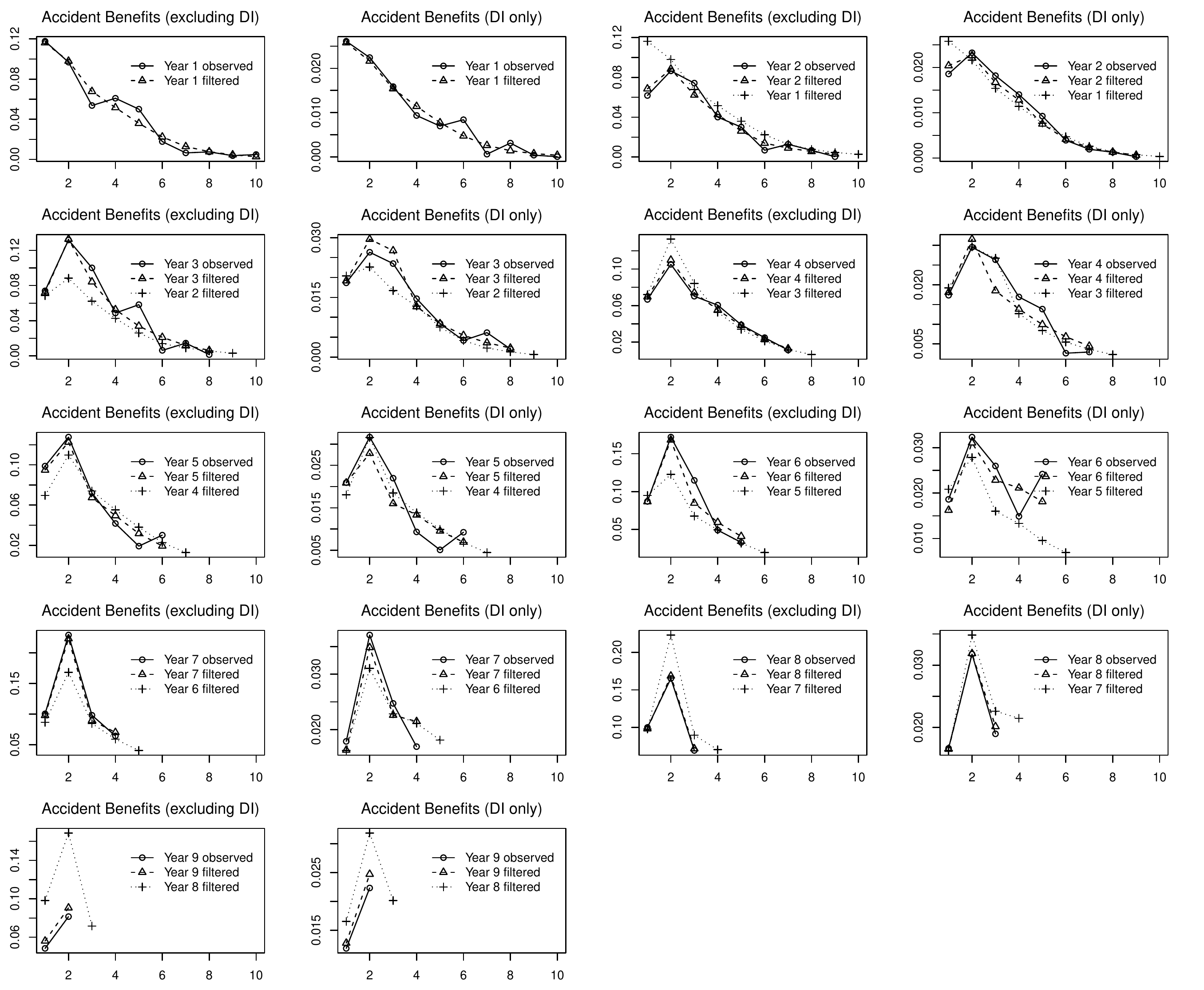}
	\caption{Tracking of claims development patterns in real data illustration}
\label{fig:tracking2}
\end{figure}

\subsection{Model used and estimation results}
A multivariate evolutionary GLM is fitted to the data set. We focus on the Tweedie sub-family of the EDF. 
For this data set, different Tweedie distributions with different power parameters $p$ are used, providing flexible dispersion modelling for the two lines. \par 

The mean structure used is  
\begin{equation}
a_i^{(n)} + r_i^{(n)}\log(j) + s_i^{(n)}j + {b_{i,1}^{(n)}\mathbbm{1}_{\lbrace j=1\rbrace} + b_{i,2}^{(n)}\mathbbm{1}_{\lbrace j=2\rbrace}} + h_t^{(n)},
\end{equation} 
where $a_i^{(n)},\,r_i^{(n)},\,s_i^{(n)},\,b_{i,1}^{(n)},\,b_{i,2}^{(n)},\,h_t^{(n)}$ are random factors that have random walk evolutions within their respective time dimension. \par

We use 50,000 samples for each time step and initialise the filter using static GLM estimates with the above mean structure. We also examine the change in claims development pattern over time and perform a number of trial runs to select somewhat informative prior distributions for parameters to reduce the degeneracy issue. The filtered (posterior) values of random factors are given in Table \ref{tab:randomestreal} and parameter estimates are given in Table \ref{sec:parestreal} in Appendix \ref{sec:evolutionaryrealdata}.

\subsection{Goodness-of-fit analysis}\label{S_Gof}
The tracking of claims development patterns is provided in Figure 
\ref{fig:tracking2}. These plots provide the observed claims development pattern for each year, the fitted patterns of that year and the previous year. Fitted patterns are calculated using filtered (posterior) estimates of random factors and parameters. These plots show that the particle filter can track changes in claim activity quite reasonably well overall, especially in the last few years. There are also quite dramatic changes in the claims development pattern in some years, for example, from year 1 to 2, year 2 to 3, which are captured well by the model. Changes within the year from year 3 to 6 are quite rapid, which are captured by the model to some extent but not fully. For years 4-6, the development pattern deviates from the shape of a Hoerl curve in the tail, and this is not captured well by the filter.\par 

Using parameter estimates in Table \ref{sec:parestreal}, the Pearson correlation coefficient between calendar year factors $h_t^{(1)}$ and $h_t^{(2)}$ is 0.3198. This is the theoretical correlation coefficient calculated using the evolution assumptions of calendar year factors in Equation \eqref{eq:calendarevolution}. The sample correlation coefficient between the filtered (posterior) values of these factors is 0.7537 with 95\% CI $(0.2362; 0.9381)$. This estimate is quite close to the coefficient of 0.6976 between calendar year residuals from the two lines in the preliminary analysis in Section \ref{sec:exploratoryanalysis}. 

\begin{figure}[htb]
	\centering
	\includegraphics[scale=0.6]{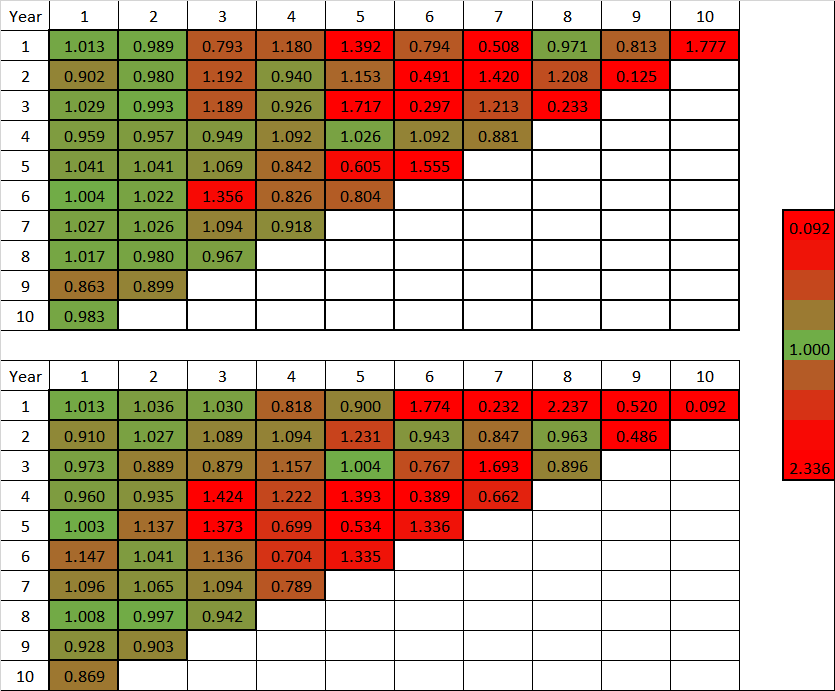}
	\caption{Heat maps of ratios of observed values to fitted values (top: Accident Benefits (excluding DI), bottom: Accident Benefits (DI only)}\label{fig:heatmapreal}
\end{figure}

We have performed a number of analyses of residuals to assess the goodness of fit. Heat maps of residuals are provided in Figure \ref{fig:heatmapreal}. The residuals in these heat maps are calculated as the ratios of observed values to fitted values. It can be observed that the goodness-of-fit is better than that of the traditional static GLM in Figure \ref{fig:evolutionaryheatcal} in Appendix \ref{sec:evolutionaryrealdata}. The goodness-of-fit is noticeably better in early development years, especially the first two years.

\begin{figure}[htb]
	\centering
	\includegraphics[scale=0.5]{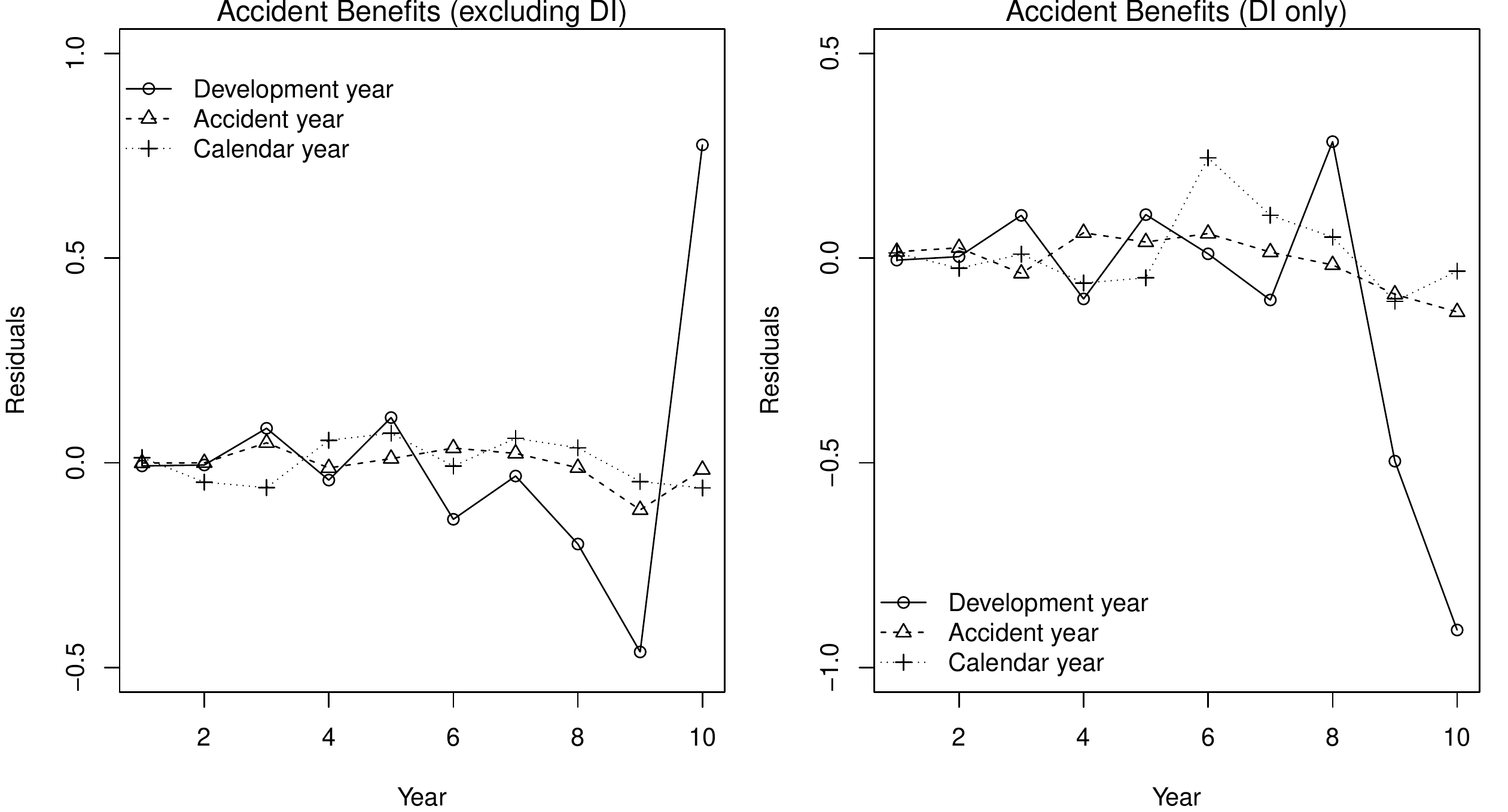}
	\caption{Plots of residuals by accident year, development year and calendar year}\label{fig:resbyyears}
\end{figure}

Plots of residuals in three dimensions: accident years, development years and calendar years are provided in Figure \ref{fig:resbyyears}. In each of these dimensions, the residual for year is calculated as the difference between the sum of observed values and the sum of fitted values for all cells in that year, divided by the sum of fitted values (to be consistent with the methodology used in Figure \ref{fig:evolutionarycalres}). The goodness-of-fit seems to deteriorate in later development years. This is due to the lack of available information for these late development lags. In addition, the use of the Hoerl curve to smooth out the whole development pattern may also contribute to this poorer fit. This is not necessarily a bad thing, as the smoothing effect of the Hoerl curve may actually be desired. Besides, observed values in these years are low, which can can also magnify the error ratios. Otherwise, the goodness-of-fit appears reasonable for accident year and calendar year residuals. \par 

There is no clear evidence of calendar year dependence in the heat maps in Figure \ref{fig:heatmapreal}. To further look for any trace of calendar year dependence, \AV{we can examine calendar year residuals for the two lines in Figure \ref{fig:resbyyears}}. The Pearson correlation coefficient of these residuals reduces to 0.1400 ($p$-value 0.6996), which is much weaker than the correlation coefficient of 0.6976 for calendar year GLM residuals in in Section \ref{sec:exploratoryanalysis} and is also insignificant. The main difference between both correlations is the capture of calendar year effects. Since the correlation is now insignificant, it seems that it was the most significant dependence driver in the data, and that our model has explained away most of the dependence in the data.

The goodness-of-fit in this illustration is not as good as the goodness-of-fit in the simulation illustration. This is expected as the synthetic data set is simulated from a theoretical model, whereas the underlying model that generates the real data set is unknown. This is to say that there may be other factors in the data that are yet to be considered and captured in the model. \par 

\begin{remark}
\AV{{The data set used in this illustration is only available in annual time interval.} When more data is available (or when data is available in a smaller interval), one could consider performing a back testing of the model to further assess its performance. However, it is also worth noting that value of reserves observed in back-testing is only one possible outcome generated by the (unknown) underlying mechanism. As a result, we can only assess the predictive power of the model using these observed reserves with a certain level of confidence.}
\end{remark}

\begin{table}[htb]
	\centering
	\begin{tabular}{crrrrrr}
		\toprule
		\hline
		\multirow{2}{*}{Year} & \multicolumn{2}{c}{AB (excluding DI)} & \multicolumn{2}{c}{AB (DI only)}& \multicolumn{2}{c}{Both lines}\\
		\cmidrule(lr){2-3}\cmidrule(lr){4-5}\cmidrule(lr){6-7}
		& \multicolumn{1}{c}{Mean} & \multicolumn{1}{c}{SD} & \multicolumn{1}{c}{Mean} & \multicolumn{1}{c}{SD} & \multicolumn{1}{c}{Mean} & \multicolumn{1}{c}{SD}  \\ 
		\hline
		1 & 187.49 & 193.56 & 38.30 & 44.31 & 225.79 & 197.85 \\ 
		2 & 579.92 & 361.71 & 216.96 & 142.92 & 796.89 & 388.72 \\ 
		3 & 1,448.14 & 698.41 & 526.54 & 299.49 & 1,974.68 & 764.69 \\ 
		4 & 2,299.10 & 1,117.06 & 940.59 & 446.71 & 3,239.70 & 1,188.88 \\ 
		5 & 5,504.30 & 2,482.17 & 3,695.25 & 1,644.89 & 9,199.55 & 2,806.42 \\ 
		6 & 13,128.14 & 5,506.59 & 5,456.64 & 2,618.54 & 18,584.78 & 6,155.75 \\ 
		7 & 17,550.89 & 8,319.55 & 6,530.34 & 3,742.70 & 24,081.23 & 9,194.94 \\ 
		8 & 19,898.82 & 11,057.94 & 7,026.08 & 6,163.77 & 26,924.90 & 12,297.21 \\ 
		9 & 31,035.09 & 20,539.99 & 14,891.23 & 24,332.31 & 45,926.32 & 32,660.82 \\ 
		\hline
		\bottomrule
	\end{tabular}
	\caption{Outstanding claims statistics by accident year}\label{tab:evolacc_forecast} 
\end{table}

\subsection{Outstanding claims forecast}
To forecast outstanding claims in the lower triangles, we utilise samples from the filtering for the upper triangles. These are samples of random factors and parameters. They are used to project future claims using the framework specification in Section \ref{sec:evolutionarymodel}. A set of simulated values of future claims can then be obtained, which is used to calculate summary statistics for the total outstanding claims liability. \AV{Future calendar year factors were simulated recursively in accordance with their random walk specification.}

\begin{figure}[htb]
	\centering
	\includegraphics[scale=0.5]{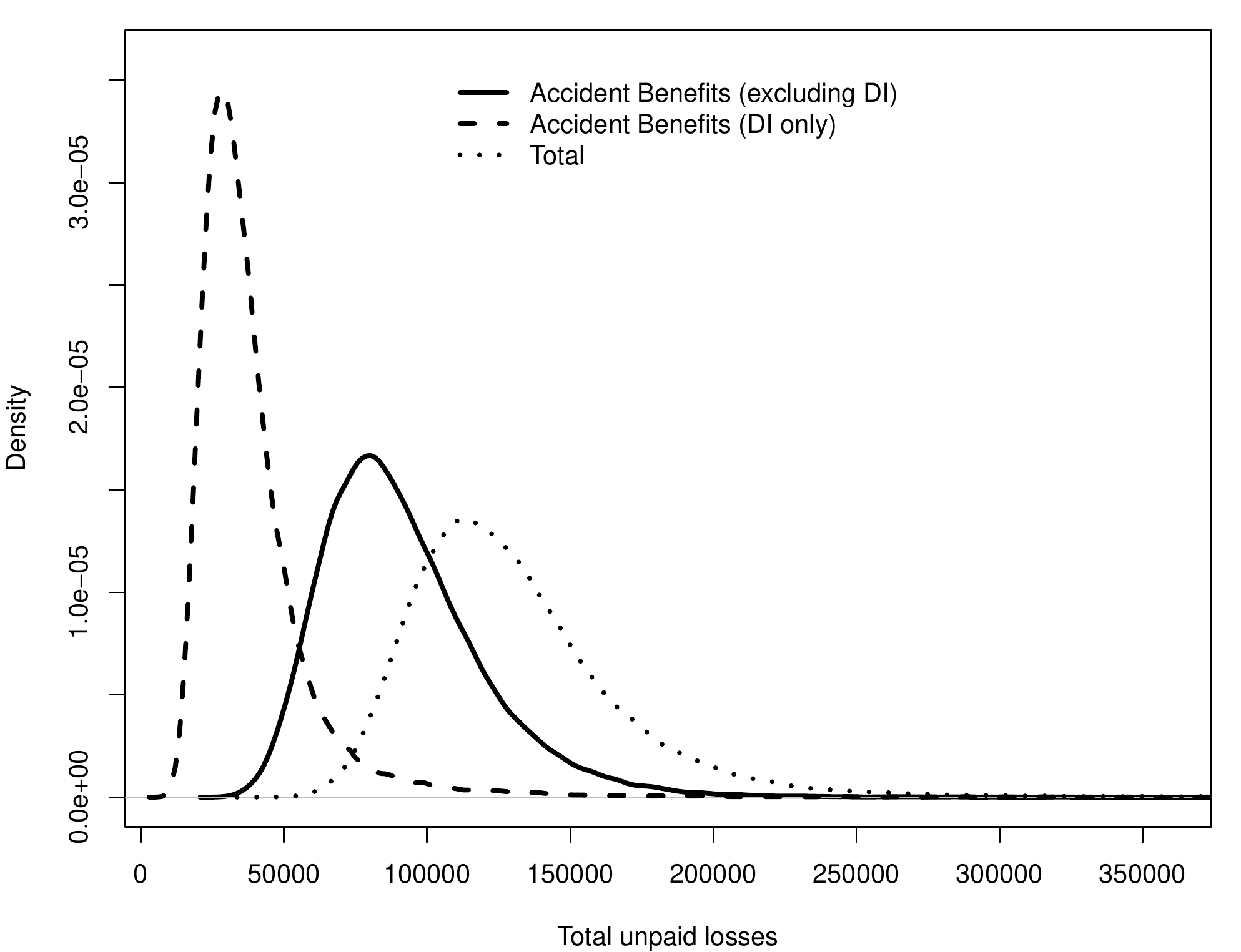}
	\caption{Kernel densities of predictive distributions of total outstanding claims in each line of business and in the aggregate portfolio}\label{Fig:evolkernel}
\end{figure}

The means and standard deviations of the total outstanding claims by accident years for each line of business and the total portfolio are summarised in Table \ref{tab:evolacc_forecast}. Summary statistics of the total outstanding claims distributions are given in Table \ref{Tab:evolsumstat} and their kernel densities are given in Figure \ref{Fig:evolkernel}. The summary statistics provided include the posterior means, standard deviations, $\text{VaR}_{75\%}$ and $\text{VaR}_{95\%}$ of the distributions of total outstanding claims for each line, as well as the total portfolio.  \par

\begin{table}[htb]
	\centering
	\begin{tabular}{lrrr}
		\toprule 
		\hline
		& \multicolumn{1}{c}{AB (excluding DI)} & \multicolumn{1}{c}{AB (DI only)} & \multicolumn{1}{c}{Both lines} \\ 
		\hline
		Mean & 91,631.90 & 39,321.94 & 130,953.84  \\ 
		SD & 28,960.41 & 26,251.88 & 40,495.76  \\ 
		$\text{VaR}_{75\%}$ & 106,307.33 & 44,375.00 & 147,631.89 \\ 
		$\text{VaR}_{95\%}$ & 143,964.73 & 75,211.60 & 198,509.08  \\ 
		\hline
		\bottomrule
	\end{tabular}
	\caption{Summary statistics of outstanding claims distributions}\label{Tab:evolsumstat}
\end{table}

The two lines, however, do not have a comonotonic dependence structure, and this allows the insurer to gain a diversification benefit when they set their risk margins. We use the following definitions of risk margins and diversification benefits (D.B.) motivated by the regulatory requirements in Australia set by the Australian Prudential Regulation Authority (and other similar regimes around the world such as in Solvency II in Europe):
\begin{equation}
{\text{Risk margin}_{\chi\%}[Y] = \max\left\lbrace \text{VaR}_{\chi\%}[Y] - E[Y];\,\dfrac{1}{2}SD[Y]\right\rbrace,}
\end{equation}
\begin{equation}
{\text{D.B.} = \dfrac{\left(\text{Risk margin}_{\chi\%}[Y_1]+\text{Risk margin}_{\chi\%}[Y_2]\right) -  \text{Risk margin}_{\chi\%}[Y_1+Y_2]}{\text{Risk margin}_{\chi\%}[Y_1]+\text{Risk margin}_{\chi\%}[Y_2]} \times 100\%}.
\end{equation}
The Risk Margin$_{75\%}$ and Risk Margin$_{95\%}$, as well as their corresponding diversification benefits are provided in Table \ref{Tab:evolriskmargin}. It can then be observed that quite significant diversification benefits can be gained as a result of allowing for (non-comonotonic) dependence across lines in the valuation of the total outstanding claims liability. 

\begin{table}[htb]
	\centering
	\begin{tabular}{lrrr|c}
		\toprule
		\hline
		& \multicolumn{1}{c}{AB (excluding DI)} & \multicolumn{1}{c}{AB (DI only)} & \multicolumn{1}{c|}{Both lines} 	& Diversification Benefit\\ 
		\hline
		Risk Margin$_{75\%}$  & 14,675.43 & 13,125.94 & 20,247.88 & 27.2\%\\ 
		Risk Margin$_{95\%}$ &   52,332.83 & 35,889.66 & 67,555.24 & 23.4\%\\ 
		\hline
		\bottomrule
	\end{tabular}
	\caption{Risk margin and diversification benefits statistics}\label{Tab:evolriskmargin}
\end{table}

\section{Discussion and practical considerations} \label{S_Discussion}

\subsection{Correlation and common shocks} \label{S_limitcorr}
There are two aspects related to the modelling of dependence between triangles, which support use of our common shock model (as opposed to relying on correlation).

In Section \ref{sec:exploratoryanalysis} we observed that the Pearson correlation coefficient between calendar year residuals was significant at 70\% when including the first year, but became insignificant when that first year was removed. This is because there was probably a structural change, but how to deal with this issue is akin to dealing with outliers. When building a model, one needs to separate the modelling of the first year (which boils down to ignoring that data as there isn't enough data to model it separately), or make a judgmental assumption about correlation. Rather than forcing extreme approaches (include the first year fully or not at all), or requiring subjective input (an assumption about correlation), our model smoothes the potential structural break naturally in an objective way: as time goes by, what happened in that first year becomes less and less influential.

Furthermore, we know from Section \ref{S_Gof} that inclusion of calendar effects in fact wipes away all correlation in the data. This further supports the idea that it is better to try and model the dependence drivers explicitly first, and use correlation as a dependence model only if and when that is not entirely successful (if there is some unexplained dependence left in the residuals); see also \citet*{AvTaWo16}. In our case, the common shock approach worked well.

\subsection{Degeneracy issues} 
The closest model to ours in the recent literature would be \citet*{Sim11}, who used particle filtering in a univariate evolutionary model. Our results appear much more satisfactory than the results for univariate evolutionary models therein. \citet*{Sim11} used the traditional sequential Monte Carlo particle filter with the variances of the disturbance terms chosen by an initial residual analysis. \citet*{Sim11} noted that the particle filter could not always keep track of the changes, and it also suffered from the degeneracy issue. The approach that we proposed, however, is an advancement of the traditional particle filter to also incorporate parameter estimation. This allows the variances of the disturbance terms to be updated upon the arrival of new observations. In addition, the particle filter used in this paper is a modification of the auxiliary particle filter. This filter typically places the re-sampling ahead of the evaluation step whereas the reverse order is typically performed in the traditional particle filter (see also the review in \citealp*{DoJo11}). The re-sampling step utilises the importance weights calculated using the look-ahead-likelihood. This aims to reduce the degeneracy issue \citep*{DoJo11,Cre12,CaGoMo07}. These could explain the better performance of our particle filter in tracking changes.

While degeneracy issues are unavoidable in particle filtering with parameter learning and particle filtering in general (see for example, \citealp*{DoGoAn00,AnDoTa05,CaJoLoPo10,Cre12}), it can be more severe for data of a higher dimension, or when uninformative priors for parameters and initial values of random factors are used \citep*{WaLiSuCo17,LiSuSaCo14}. \AV{However, it is noted that degeneracy appears to affect mainly the random parameters of the model in the simulation illustration of Section \ref{sec:evolutionarysim}.  Confidence intervals of the parameters there are small to the point where they are essentially fixed parameters.  While recognition of their randomness in confidence intervals would be preferable, it should be noted that the end result is of a quality equal to that of a model that specifies these as (almost) fixed parameters.  There is no sign of particle degeneracy in relation to the actual fixed parameters of the model.}

\subsection{Filtering based on accident years}

The particle filter and the dual Kalman filter that we introduce in this chapter are accident-year-based. It means that they update estimates of random factors and parameters from one accident year to another. This is to utilise the greater availability of data in the first accident year, which helps initialise the filters more accurately. However, given that new information typically arrives by calendar years, it can be better for risk management purposes to have the filters run by calendar years. This, however, creates additional complications due to the lack of data for initialisation. In addition, accident year factors and development year factors typically evolve by rows. This can create challenges for calendar-year-based filters. In this paper we consider a special treatment for calendar year factors in our accident-year-based filter. A similar treatment could be considered for accident year and development year factors in the development of calendar-based-filters.

\subsection{Compensation due to collinearity of the factors}

Compensation can occur across estimates of random factors where one factor is consistently overestimated and this is offset by an underestimation of another factor. This issue can arise in any model that incorporates all three factors, accident year factor, development year factor and calendar year factor due to the collinearity between them. The mis-allocations between these factors can largely offset to give an overall reasonable fit. However, caution needs to be taken in cases where changes in claim experience are sudden and some reasonableness checks can be useful.

\subsection{Backsmoothing}

It is often desirable to perform back smoothing following filtering to obtain estimates using all available information. Back smoothing is a lot easier to accomplish for Gaussian models because of the availability of the estimates in closed-form. For non-Gaussian models, particle smoothing can be used. However, it is not an easy task due to particle degeneracy \citep*{DoJo11}. This problem is further escalated for the particle filtering with parameter learning algorithm in Section \ref{sec:evolutionarygenest} as parameters are also incorporated in the on-line estimation. We do not perform back smoothing for the evolutionary GLM framework. Future research could further investigate this aspect, especially with regard to addressing the degeneracy problem.

\subsection{\AV{Size of triangles and calibration}}
\AV{To assess the size of triangles required for a good calibration, we have run the filter on the simulated data in Section \ref{sec:evolutionarysim}, using the sub-triangles of dimension 6, 9 and 12 respectively. The results are then compared with those from the full triangle. }

\AV{The above study shows that the resultant fitted observations can track the actual observations quite closely even when triangles of a small size are used. It is worth noting that even though random factors $r_i^{(n)}$ and $s_i^{(n)}$ for an undeveloped accident year $i$ can be calibrated reasonably well, the accuracy of estimation would improve as more data is observed for accident year $i$. This is because these two factors are used to formulate a Hoerl curve which captures the total claims development pattern in accident year $i$. The number of observations required for reliable estimation would be likely to depend on the length of the payment tail.}

\section{Conclusion}\label{sec:evolutionaryremarks}
Insurers typically experience changes in their claim experience over time, making the application of static models with deterministic parameters no longer straightforward. We capture this common data feature in a multivariate evolutionary GLM framework. This framework utilises the very popular and rich GLM structure, hence provides great flexibility in modelling. We extend the traditional GLM framework in loss reserving on two fronts. Firstly, we allow parameters of the traditional GLM framework to evolve, and are hence enabling changes in claim experience to be captured naturally in an elegant manner. This helps provide a clear picture of the historical experience. Secondly, we introduce dependence across lines using a common shock approach with an explicit and easy-to-interpret dependence structure.  We specifically target the calendar year dependence in the specification of our framework. This, however, can be modified easily to incorporate other sources of dependence such as common accident year effects, and common development year effects. The framework introduced also specifies the use of random walks for the evolution of random factors for the sake of simplification. This could be extended with a more complex time series model, but this might come at a cost of more extensive parametrisation.

Together with the development of the multivariate evolutionary GLM framework, we also contribute to the literature with the formulation of two adaptive estimation approaches: a particle filter with parameter learning for the general framework, and a dual Kalman filter for the special case of Gaussian models. These filters are real-time devices that recursively provide posterior estimates of random factors and parameters upon the arrival of new information. They give more weight to more recent data, which increases the likelihood of a more accurate projection of future claims. In the special structure of reserving data with two different time dimensions, the application of a typical adaptive estimation approach is not straightforward. We take into account this difficulty in the development of the filters. Specifically, calendar year factors are treated as fixed effect in the period where they do not evolve as we consider the evolution in the accident year dimension. 

Theoretical developments are demonstrated using illustrations on a simulated data set and a real data set. The results show that factor estimates can adapt to changes in claim experience reasonably well. Some practical considerations are also raised through these illustrations. A common issue with particle filters is particle degeneracy where the number of samples with non-negligible weight drops significantly after a few time periods. A mis-allocation across estimates of factors can also be observed in models which consider all three factors, accident year, development year and calendar year factors due to the collinearity between them. A careful selection of priors and initial value and frequent reasonableness check may be required to mitigate these issues. \AV{Future research could look into a new development of particle filtering for adaptive reserving that can overcome the issues of particle degeneracy.}

\section*{R codes}
Full R codes for Sections \ref{sec:evolutionarysim} and \ref{sec:evolutionaryreal} are available upon request by contacting the authors.

\section*{Acknowledgements}
Results in this paper were presented at a research seminar at UNSW in Sydney (Australia), at The Australasian Actuarial Education and Research Symposium in 2018, at the $L^2$ Seminars in Lausanne (Switzerland) \AV{and at the $23^{rd}$ International Congress on Insurance: Mathematics and Economics in Munich (Germany)} in 2019. The authors are grateful for constructive comments received from colleagues who attended those presentations. \AV{The authors are also thankful to the two anonymous reviewers for  constructive comments that helped improve the paper.} 

This research was  supported under Australian Research Council's Linkage (LP130100723, with funding partners Allianz Australia Insurance Ltd, Insurance Australia Group Ltd, and Suncorp Metway Ltd) and Discovery (DP200101859) Projects funding schemes. Furthermore, Phuong Anh Vu acknowledges financial support from a University International Postgraduate Award/University Postgraduate Award and supplementary scholarships provided by the UNSW Business School. The views expressed herein are those of the authors and are not necessarily those of the supporting organisations.

\section*{References}

\bibliographystyle{elsarticle-harv}
\bibliography{libraries}

\appendix

	\section{Appendices}\label{sec:evolutionaryappendix}

	\subsection{Simulated data set and estimation results}\label{sec:evolutionarysimdata2}
	\begin{landscape}        
		\begin{table}[H]
			\centering\resizebox{1.2\textheight}{!}{
				\begin{tabular}{crrrrrrrrrrrrrrr}
					\toprule
					\hline
					Year	& 1 & 2 & 3 & 4 & 5 & 6 & 7 & 8 & 9 & 10 & 11 & 12 & 13 & 14 & 15 \\ 
					\hline

						1 & 791.47 & 828.39 & 555.17 & 408.36 & 220.69 & 144.34 & 81.33 & 38.77 & 19.39 & 19.85 & 4.95 & 3.69 & 0.06 & 0.56 & 0.00 \\ 
						2 & 851.29 & 946.12 & 706.62 & 435.07 & 296.40 & 141.62 & 84.15 & 45.51 & 33.97 & 13.45 & 6.45 & 1.94 & 1.73 & 1.47 &  \\ 
						3 & 688.81 & 742.36 & 633.17 & 354.99 & 230.12 & 140.81 & 76.95 & 37.40 & 20.79 & 17.41 & 4.16 & 2.25 & 1.30 &  &  \\ 
						4 & 743.39 & 728.81 & 530.71 & 304.13 & 168.36 & 81.24 & 61.89 & 33.15 & 22.10 & 16.19 & 2.19 & 3.31 &  &  &  \\ 
						5 & 674.33 & 562.87 & 463.11 & 290.24 & 176.52 & 89.16 & 46.72 & 27.53 & 14.82 & 7.97 & 2.71 &  &  &  &  \\ 
						6 & 618.23 & 579.59 & 409.98 & 268.87 & 154.98 & 83.93 & 40.90 & 23.45 & 10.05 & 2.52 &  &  &  &  &  \\ 
						7 & 587.55 & 504.90 & 357.14 & 190.76 & 103.46 & 60.88 & 36.08 & 20.18 & 6.75 &  &  &  &  &  &  \\ 
						8 & 629.49 & 589.69 & 439.96 & 261.07 & 143.96 & 74.47 & 39.26 & 22.72 &  &  &  &  &  &  &  \\ 
						9 & 702.89 & 678.30 & 488.67 & 315.59 & 219.22 & 118.56 & 58.06 &  &  &  &  &  &  &  &  \\ 
						10 & 942.99 & 978.78 & 655.10 & 448.96 & 215.67 & 174.70 &  &  &  &  &  &  &  &  &  \\ 
						11 & 1,101.07 & 1,310.81 & 975.22 & 641.96 & 417.15 &  &  &  &  &  &  &  &  &  &  \\ 
						12 & 1,075.32 & 1,188.63 & 1,036.26 & 672.44 &  &  &  &  &  &  &  &  &  &  &  \\ 
						13 & 1,027.16 & 1,459.89 & 1,261.43 &  &  &  &  &  &  &  &  &  &  &  &  \\ 
						14 & 926.01 & 1,213.62 &  &  &  &  &  &  &  &  &  &  &  &  &  \\ 
						15 & 861.12 &  &  &  &  &  &  &  &  &  &  &  &  &  &  \\ 
					\hline 
					\bottomrule
			\end{tabular}}
			\caption{Simulated triangle 1 }\label{tab:evolutionarysimdata21}
		\end{table}
		
		\begin{table}[H]
			\centering
			\resizebox{1.35\textheight}{!}{
				\begin{tabular}{crrrrrrrrrrrrrrr}
					\toprule
					\hline
					Year	& 1 & 2 & 3 & 4 & 5 & 6 & 7 & 8 & 9 & 10 & 11 & 12 & 13 & 14 & 15 \\ 
					\hline
					1 & 1,314.30 & 3,392.07 & 4,921.77 & 5,510.09 & 5,586.52 & 5,298.04 & 4,476.64 & 3,895.52 & 3,445.22 & 2,740.11 & 1,961.77 & 1,542.13 & 1,286.08 & 889.59 & 635.24 \\ 
					2 & 1,131.31 & 3,101.68 & 4,246.48 & 4,868.79 & 4,822.71 & 4,609.38 & 3,775.83 & 3,308.23 & 2,674.09 & 1,950.31 & 1,736.38 & 1,192.11 & 840.96 & 723.24 &  \\ 
					3 & 1,148.99 & 3,144.70 & 4,278.32 & 4,378.19 & 4,557.47 & 4,316.18 & 3,905.05 & 3,180.89 & 2,797.56 & 2,129.81 & 1,672.20 & 1,396.30 & 995.19 &  &  \\ 
					4 & 876.20 & 2,417.08 & 3,073.04 & 3,739.61 & 3,380.83 & 3,014.43 & 2,650.25 & 2,182.04 & 1,649.63 & 1,248.75 & 874.43 & 713.64 &  &  &  \\ 
					5 & 961.52 & 2,242.44 & 2,908.57 & 3,161.07 & 3,221.94 & 2,786.73 & 2,444.58 & 1,949.89 & 1,459.16 & 1,156.00 & 853.76 &  &  &  &  \\ 
					6 & 732.80 & 2,271.59 & 3,222.15 & 3,396.90 & 3,444.96 & 3,399.86 & 3,097.23 & 2,522.98 & 1,851.79 & 1,471.42 &  &  &  &  &  \\ 
					7 & 732.67 & 1,940.64 & 3,255.24 & 4,060.41 & 4,382.26 & 4,085.70 & 3,654.16 & 3,350.06 & 2,758.58 &  &  &  &  &  &  \\ 
					8 & 785.16 & 2,148.23 & 3,257.28 & 3,734.23 & 3,748.20 & 3,656.86 & 3,148.13 & 3,055.73 &  &  &  &  &  &  &  \\ 
					9 & 711.79 & 2,285.66 & 3,304.61 & 3,876.34 & 4,036.46 & 3,519.46 & 3,601.98 &  &  &  &  &  &  &  &  \\ 
					10 & 708.64 & 2,140.87 & 3,145.20 & 3,304.59 & 4,009.19 & 3,519.30 &  &  &  &  &  &  &  &  &  \\ 
					11 & 803.85 & 2,026.72 & 3,314.95 & 3,882.90 & 4,377.03 &  &  &  &  &  &  &  &  &  &  \\ 
					12 & 816.20 & 1,863.28 & 3,039.58 & 3,645.57 &  &  &  &  &  &  &  &  &  &  &  \\ 
					13 & 602.57 & 2,037.97 & 2,960.59 &  &  &  &  &  &  &  &  &  &  &  &  \\ 
					14 & 708.14 & 2,237.89 &  &  &  &  &  &  &  &  &  &  &  &  &  \\ 
					15 & 728.01 &  &  &  &  &  &  &  &  &  &  &  &  &  &  \\ 
					\hline
					\bottomrule
			\end{tabular}}
			\caption{Simulated triangle 2}\label{tab:evolutionarysimdata22}
		\end{table}
	\end{landscape}
	\begin{table}[ht]
		\centering
		\begin{tabular}{c|c|cc|cc|cc|cc}
			\toprule
			\hline
			n	& i	&  \multicolumn{2}{c|}{$a_{i}^{(n)}$} & \multicolumn{2}{c|}{$r_{i}^{(n)}$} & \multicolumn{2}{c|}{$s_{i}^{(n)}$} & \multicolumn{2}{c}{$h_{i}^{(n)}$}\\
			& 	&  True & Estimate & True& Estimate& True & Estimate & True & Estimate\\
			\hline
			\multirow{15}{0.3cm}{$(1)$}&1 & 6.9111 & 6.9772 & 1.2867 & 1.1637 & -0.8014 & -0.7669 & 0.5000 & 0.4500 \\ 
		&	2 & 7.1203 & 7.0228 & 1.1831 & 1.2624 & -0.7783 & -0.7905 & 0.5025 & 0.4583 \\ 
		&	3 & 6.8500 & 6.8797 & 1.2921 & 1.2786 & -0.7982 & -0.7882 & 0.4974 & 0.4569 \\ 
		&	4 & 6.7755 & 6.8645 & 1.2128 & 1.1614 & -0.7977 & -0.7849 & 0.4985 & 0.4503 \\ 
		&	5 & 6.7846 & 6.7775 & 1.1308 & 1.1068 & -0.7965 & -0.7809 & 0.4869 & 0.4636 \\ 
		&	6 & 6.7512 & 6.7561 & 1.1458 & 1.1150 & -0.8197 & -0.8041 & 0.4824 & 0.4706 \\ 
		&	7 & 6.6737 & 6.6867 & 1.0900 & 1.0812 & -0.8394 & -0.8245 & 0.4900 & 0.4667 \\ 
		&	8 & 6.6937 & 6.8043 & 1.0746 & 1.0563 & -0.7904 & -0.7970 & 0.4872 & 0.4399 \\ 
		&	9 & 6.8156 & 6.8286 & 1.1102 & 1.0903 & -0.7663 & -0.7491 & 0.4925 & 0.4729 \\ 
		&	10 & 7.0346 & 7.1467 & 1.1100 & 1.0572 & -0.7412 & -0.7471 & 0.4923 & 0.4398 \\ 
		&	11 & 7.1906 & 7.2977 & 1.2651 & 1.1593 & -0.7347 & -0.7133 & 0.5041 & 0.4394 \\ 
		&	12 & 7.1507 & 7.2521 & 1.3664 & 1.1840 & -0.7553 & -0.7031 & 0.5028 & 0.4336 \\ 
		&	13 & 7.1222 & 7.2113 & 1.4539 & 1.3373 & -0.6913 & -0.6700 & 0.4995 & 0.4447 \\ 
		&	14 & 7.0539 & 7.0465 & 1.5131 & 1.3268 & -0.7400 & -0.6537 & 0.4948 & 0.4600 \\ 
		&	15 & 6.9438 & 6.9839 & 1.4693 & 1.3363 & -0.6674 & -0.6564 & 0.4973 & 0.4403 \\ 
			\hline
			\multirow{15}{0.3cm}{$(2)$}&1 & 7.0908 & 7.1637 & 2.0212 & 1.9650 & -0.4343 & -0.4275 & \AV{0.5000} & 0.4500 \\ 
				&	2 & 6.9679 & 7.0941 & 2.0064 & 1.9393 & -0.4422 & -0.4376 & \AV{0.5041} & 0.4533 \\ 
				&	3 & 6.9701 & 7.0290 & 1.9750 & 1.9282 & -0.4368 & -0.4256 & \AV{0.5117} & 0.4523 \\ 
				&	4 & 6.8030 & 6.9095 & 1.9991 & 1.9200 & -0.4769 & -0.4635 & \AV{0.5046} & 0.4666 \\ 
				&	5 & 6.7276 & 6.8444 & 1.9401 & 1.9091 & -0.4614 & -0.4644 & \AV{0.4993} & 0.4679 \\ 
				&	6 & 6.6243 & 6.7794 & 1.9961 & 1.9261 & -0.4351 & -0.4328 & \AV{0.4866} & 0.4651 \\ 
				&	7 & 6.5513 & 6.6753 & 2.0254 & 2.0316 & -0.3946 & -0.4090 & \AV{0.4991} & 0.4835 \\ 
				&	8 & 6.5454 & 6.6182 & 2.0025 & 2.0224 & -0.4018 & -0.4114 & \AV{0.4860} & 0.4748 \\ 
				&	9 & 6.5506 & 6.6417 & 2.0265 & 2.0118 & -0.3959 & -0.4076 & \AV{0.4883} & 0.4704 \\ 
				&	10 & 6.5493 & 6.5592 & 2.0127 & 2.0912 & -0.4040 & -0.4261 & \AV{0.4880} & 0.4647 \\ 
				&	11 & 6.5508 & 6.5689 & 1.9884 & 2.0683 & -0.3751 & -0.3991 & \AV{0.4911} & 0.4570 \\ 
				&	12 & 6.4854 & 6.5595 & 2.0249 & 2.0520 & -0.4015 & -0.4111 & \AV{0.4872} & 0.4364 \\ 
				&	13 & 6.4746 & 6.5175 & 2.1207 & 2.0716 & -0.4055 & -0.4130 & \AV{0.4843} & 0.4520 \\ 
				&	14 & 6.5039 & 6.5914 & 2.1752 & 2.0839 & -0.3687 & -0.4179 & \AV{0.4760} & 0.4413 \\ 
				&	15 & 6.4701 & 6.5787 & 2.2322 & 2.0846 & -0.3650 & -0.4203 & \AV{0.4736} & 0.4515 \\ 
			\hline
			\bottomrule
		\end{tabular}
		
		\caption{Filtered (posterior) estimates of random factors }\label{tab:randomestsim2}
	\end{table}
	\begin{table}[H]
		\centering
		\begin{tabular}{cccc}
			\toprule
			\hline
			& True value & Estimate & 90\% CI\\ 
			\hline
			$\phi^{(1)}$ & 0.4000 & 0.4730 & (0.4724; 0.4743) \\ 
			&&&\\[-1em]
			$\sigma^2_{{}_a\epsilon^{(1)}}$ & 0.0100 & 0.0180& (0.0178; 0.0181)  \\  
			&&&\\[-1em]
			$\sigma^2_{{}_r\epsilon^{(1)}}$ &0.0050 & 0.0078 & (0.0078; 0.0078)  \\ 
			&&&\\[-1em]
			$\sigma^2_{{}_s\epsilon^{(1)}}$ & 0.0010 & 0.0014 & (0.0014; 0.0014) \\ 
			&&&\\[-1em]
			$\sigma^2_{{}_h\epsilon^{(1)}}$ &0.0050 & 0.0047 & (0.0047; 0.0047) \\ 
			&&&\\[-1em]
			$\lambda^{(1)}$ & 0.6000 & 0.6752 & (0.6750; 0.6755) \\ 
			&&&\\[-1em]
			$p^{(1)}$ &1.2700 & 1.2844 & (1.2843; 1.2845) \\ 
			\hline
			$\phi^{(2)}$ &  0.5000 & 0.3954 & (0.3939; 0.3984) \\ 
			&&&\\[-1em]
			$\sigma^2_{{}_a\epsilon^{(2)}}$ &0.0050 & 0.0057 & (0.0057; 0.0057) \\ 
			&&&\\[-1em]
			$\sigma^2_{{}_r\epsilon^{(2)}}$ & 0.0020 & 0.0010 & (0.0010; 0.0010) \\ 
			&&&\\[-1em]
			$\sigma^2_{{}_s\epsilon^{(2)}}$ &0.0005 & 0.0006 & (0.0006; 0.0006) \\ 
			&&&\\[-1em]
			$\sigma^2_{{}_h\epsilon^{(2)}}$ &0.0050 & 0.0032 & (0.0032; 0.0032) \\ 
			&&&\\[-1em]
			$\lambda^{(2)}$ &0.8000 & 0.8281 & (0.8279; 0.8283) \\ 
			$p^{(2)}$ & 1.3500 & 1.4811 & (1.4810; 1.4813) \\ 
			&&&\\[-1em]
			$\sigma^2_{{}_h\tilde{\epsilon}}$ &   0.0050 & 0.0063 & (0.0063; 0.0063) \\ 
			&&&\\[-1em]
			\hline 
			\bottomrule
		\end{tabular}
		\caption{Posterior estimates of parameters}\label{tab:evolutionarysimest2}
	\end{table}
	\subsection{Empirical data set and estimation results}\label{sec:evolutionaryrealdata}
	This data set is drawn from \citet*{CoGeAb16}.
	\begin{landscape}
		\begin{table}[H]
			\centering                      
			\begin{tabular}{crrrrrrrrrrr}
				\toprule
				\hline
				Year & \multicolumn{1}{c}{Premium} &  \multicolumn{1}{c}{1}     & \multicolumn{1}{c}{2}     & \multicolumn{1}{c}{3}     & \multicolumn{1}{c}{4}     & \multicolumn{1}{c}{5 }    & \multicolumn{1}{c}{6 }    & \multicolumn{1}{c}{7 }    & \multicolumn{1}{c}{8}     & \multicolumn{1}{c}{9 } & \multicolumn{1}{c}{10}\\
				\hline
				1 & 116,491 & 13,714 & 24,996 & 31,253 & 38,352 & 44,185 & 46,258 & 47,019 & 47,894 & 48,334 & 48,902 \\ 
				2 & 111,467 & 6883 & 16,525 & 24,796 & 29,263 & 32,619 & 33,383 & 34,815 & 35,569 & 35,612 &  \\ 
				3 & 107,241 & 7933 & 22,067 & 32,801 & 38,028 & 44,274 & 44,948 & 46,507 & 46,665 &  &  \\ 
				4 & 105,687 & 7052 & 18,166 & 25,589 & 31,976 & 36,092 & 38,720 & 39,914 &  &  &  \\ 
				5 & 105,923 & 10,463 & 23,982 & 31,621 & 36,039 & 38,070 & 41,260 &  &  &  &  \\ 
				6 & 111,487 & 9697 & 28,878 & 41,678 & 47,135 & 50,788 &  &  &  &  &  \\ 
				7 & 113,268 & 11,387 & 37,333 & 48,452 & 55,757 &  &  &  &  &  &  \\ 
				8 & 121,606 & 12,150 & 32,250 & 40,677 &  &  &  &  &  &  &  \\ 
				9 & 110,610 & 5348 & 14,357 &  &  &  &  &  &  &  &  \\ 
				10 & 104,304 & 4,612 &  &  &  &  &  &  &  &  &  \\ 
				\hline
				\bottomrule
			\end{tabular}
			\caption{Accident Benefits excluding Disability Income (cumulative claims)}\label{Tab:evolutionaryreal1}
		\end{table}
		
		\begin{table}[H]
			\centering
			\begin{tabular}{crrrrrrrrrrr}
				\toprule
				\hline
				Year & \multicolumn{1}{c}{Premium} &  \multicolumn{1}{c}{1}     & \multicolumn{1}{c}{2}     & \multicolumn{1}{c}{3}     & \multicolumn{1}{c}{4}     & \multicolumn{1}{c}{5 }    & \multicolumn{1}{c}{6 }    & \multicolumn{1}{c}{7 }    & \multicolumn{1}{c}{8}     & \multicolumn{1}{c}{9 } & \multicolumn{1}{c}{10}\\
				\hline
				1 & 116,491 & 3,043 & 5,656 & 7,505 & 8,593 & 9,403 & 10,380 & 10,450 & 10,812 & 10,856 & 10,860 \\ 
				2 & 111,467 & 2,070 & 4662 & 6,690 & 8,253 & 9,286 & 9,724 & 9,942 & 10,086 & 10,121 &  \\ 
				3 & 107,241 & 2,001 & 4,825 & 7,344 & 8,918 & 9,824 & 10,274 & 10,934 & 11,155 &  &  \\ 
				4 & 105,687 & 1,833 & 4,953 & 7,737 & 9,524 & 10,986 & 11,267 & 11,579 &  &  &  \\ 
				5 & 105,923 & 2,217 & 5,570 & 7,898 & 8,885 & 9,424 & 10,402 &  &  &  &  \\ 
				6 & 111,487 & 2,076 & 5,681 & 8,577 & 10,237 & 12,934 &  &  &  &  &  \\ 
				7 & 113,268 & 2,025 & 6,225 & 9,027 & 10,945 &  &  &  &  &  &  \\ 
				8 & 121,606 & 2,024 & 5,888 & 8,196 &  &  &  &  &  &  &  \\ 
				9 & 110,610 & 1,311 & 3,780 &  &  &  &  &  &  &  &  \\ 
				10 & 104,304 &  912 &  &  &  &  &  &  &  &  &  \\ 
				\hline
				\bottomrule
			\end{tabular}
			\caption{Accident Benefits - Disability Income only (cumulative claims)}\label{Tab:evolutionaryreal2}
		\end{table}
	\end{landscape}
\begin{figure}[H]
	\begin{center}
		\includegraphics[scale=0.6]{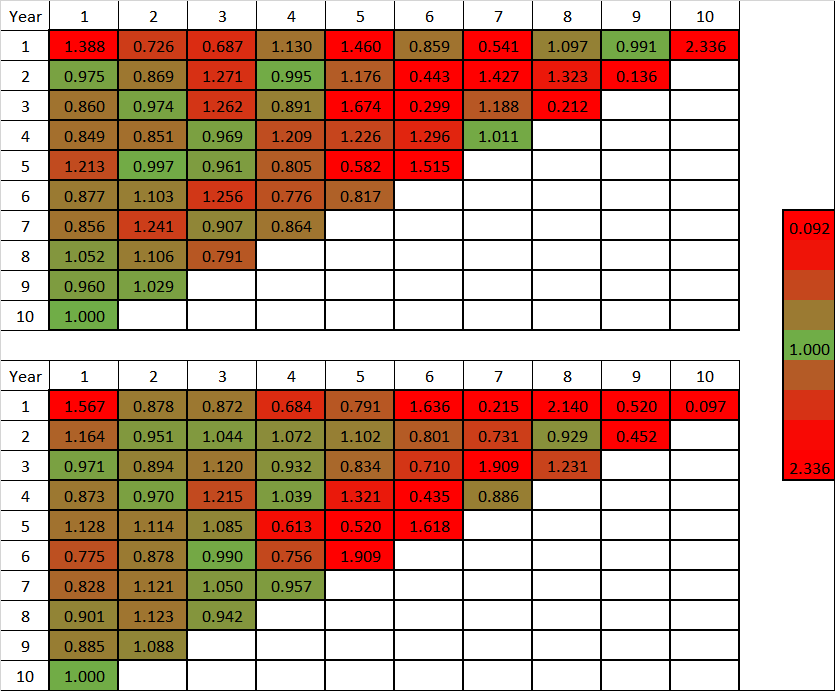}
	\end{center}
	\caption{Heat maps of ratios of observed values to GLM fitted values (top: Accident Benefits (excluding DI), bottom: Accident Benefits (DI only)}\label{fig:evolutionaryheatcal}
\end{figure}
	\begin{table}[H]
		\centering
		\begin{tabular}{c|c|cccccc}
			\toprule
			\hline
			$n$	& $i$	&  $a_i^{(n)}$ & $r_i^{(n)}$ & $s_i^{(n)}$ & $b_{i,1}^{(n)}$ & $b_{i,2}^{(n)}$& $h_i^{(n)}$\\
			\hline
			\multirow{15}{0.3cm}{$(1)$}& 1 & -2.0093 & 1.3115 & -0.6984 & 0.5553 & 0.2005 & 0.0000 \\
			&	2 & -2.0142 & 1.2026 & -0.7005 & 0.0362 & 0.1557 & -0.0028 \\ 
			&	3 & -2.0479 & 1.9093 & -0.8420 & 0.2509 & 0.2237 & 0.0068 \\ 
			&	4 & -2.0474 & 1.9407 & -0.8349 & 0.0359 & 0.1674 & 0.1814 \\ 
			&	5 & -2.0153 & 1.9509 & -0.8733 & 0.5484 & 0.4951 & -0.0153 \\ 
			&	6 & -1.8899 & 2.1597 & -0.9237 & 0.5494 & 0.6518 & -0.1817 \\ 
			&	7 & -1.9436 & 2.1534 & -0.8723 & 0.6715 & 0.8886 & -0.1791 \\ 
			&	8 & -1.9963 & 2.0434 & -0.8846 & 0.7631 & 0.7925 & -0.2023 \\ 
			&	9 & -2.0942 & 1.8848 & -0.8674 & 0.2985 & 0.3903 & -0.2185 \\ 
			&	10 & -2.0660 & 1.9685 & -0.9101 & 0.1290 & 0.3581 & -0.2548 \\ 
			\hline
			\multirow{15}{0.3cm}{$(2)$}&	 1 & -3.6999 & 1.9821 & -0.8779 & 0.9202 & 0.2526 & 0.0000 \\ 
			&	2 & -3.6618 & 2.1940 & -0.9283 & 0.6658 & 0.1919 & 0.0322 \\ 
			&	3 & -3.6864 & 2.0891 & -0.8281 & 0.6168 & 0.2896 & -0.0561 \\ 
			&	4 & -3.6632 & 2.2922 & -0.8690 & 0.4334 & 0.1267 & 0.0855 \\ 
			&	5 & -3.6485 & 2.1814 & -0.8517 & 0.3686 & 0.4682 & 0.2618 \\ 
			&	6 & -3.6084 & 2.1829 & -0.7584 & 0.4518 & 0.4729 & -0.2053 \\ 
			&	7 & -3.6036 & 2.0957 & -0.7528 & 0.5783 & 0.5821 & -0.3378 \\ 
			&	8 & -3.6202 & 1.9516 & -0.7664 & 0.5515 & 0.5776 & -0.2683 \\ 
			&	9 & -3.5940 & 1.9978 & -0.8600 & 0.3277 & 0.3639 & -0.2342 \\ 
			&	10 & -3.6191 & 2.0065 & -0.8329 & -0.0011 & 0.2890 & -0.1454 \\ 
			\hline
			\bottomrule
		\end{tabular}
		
		\caption{Filtered (posterior) estimates of random factors}\label{tab:randomestreal}
	\end{table}
\begin{table}[H]
	\centering
	\begin{tabular}{c|cc|cc}
		\toprule
		\hline 
		& \multicolumn{2}{c|}{$n=1$} &  \multicolumn{2}{c}{$n=2$}\\
		\hline
		& Estimate & 90\% CI & Estimate & 90\% CI \\ 
		\hline
		$\phi^{(n)}$ & 0.0069 & (0.0063; 0.0073) & 0.0071 & (0.0063; 0.0081) \\  
		&&&&\\[-1em]
		$p^{(n)}$ & 1.2537 & (1.2417; 1.2686) & 1.3602 & (1.3495; 1.3693) \\ 
		&&&&\\[-1em]
		\hline
		$\sigma^2_{{}_a\epsilon^{(n)}}$ & 0.0048 & (0.0043; 0.0053) & 0.0035 & (0.0032; 0.0040) \\ 
		&&&&\\[-1em]
		$\sigma^2_{{}_r\epsilon^{(n)}}$ & 0.0499 & (0.0478; 0.0521) & 0.0788 & (0.0752; 0.0846) \\ 
		&&&&\\[-1em]
		$\sigma^2_{{}_s\epsilon^{(n)}}$ & 0.0078 & (0.0074; 0.0082) & 0.0156 & (0.0138; 0.0178) \\ 
		&&&&\\[-1em]
		$\sigma^2_{{}_{b_{,1}}\epsilon^{(n)}}$ & 0.2055 & (0.2005; 0.2118) & 0.2047 & (0.1958; 0.2124) \\ 
		&&&&\\[-1em]
		$\sigma^2_{{}_{b_{,2}}\epsilon^{(n)}}$ & 0.1315 & (0.1272; 0.1359) & 0.1181 & (0.1039; 0.1301) \\ 
		&&&&\\[-1em]
		\hline
		$\sigma^2_{{}_h\epsilon^{(n)}}$ & 0.0753 & (0.0725; 0.0789) & 0.1119 & (0.0966; 0.1356) \\ 
		&&&&\\[-1em]
		$\lambda^{(n)}$ & 0.7216 & (0.6730; 0.7808) & 0.6761 & (0.6223; 0.7488) \\ 
		&&&&\\[-1em]
		$\sigma^2_{{}_h\tilde{\epsilon}}$ &  0.1123 & (0.1057; 0.1200) &  &   \\ 
		&&&&\\
		\hline
		\bottomrule
	\end{tabular}
	\caption{Posterior estimates of parameters}\label{sec:parestreal}
\end{table}

\end{document}